\documentclass{emulateapj} 
\usepackage{apjfonts}              
\usepackage{graphicx}
\usepackage{amsmath}

\graphicspath{{./}}

\newcommand{\Msun}{M_{\odot}}

\newcommand{\Lsun}{L_{\odot}}
\newcommand{\Lbol}{L_{\rm {bol}}}
\newcommand{\LB}{L_{\rm B}}
\newcommand{\Lfir}{L_{\rm {FIR}}}
\newcommand{\Ledd}{L_{\rm {Edd}}}

\newcommand{\MBH}{M_{\rm{BH}}}
\newcommand{\Mbulge}{M_{\rm{bulge}}}
\newcommand{\Mstar}{M_{\rm{star}}}
\newcommand{\msigma}{\MBH\rm{-}\sigma}
\newcommand{\mbulge}{\rm{\MBH\rm{-}\Mbulge}}
\newcommand{\zquasar}{J1148+5251}
\newcommand{\mdm}{m_{\rm{dm}}}

\newcommand{\Gpc}{\rm {Gpc}}
\newcommand{\Mpc}{\rm {Mpc}}
\newcommand{\kpc}{\rm {kpc}}
\newcommand{\yr}{\rm {yr}}

\newcommand{\Mvir}{M_{\rm{vir}}}
\newcommand{\Vvir}{V_{\rm{vir}}}
\newcommand{\Rvir}{R_{\rm{vir}}}
\newcommand{\Cvir}{C_{\rm{vir}}}
\newcommand{\rhoc}{\rho_{\rm{crit}}}

\def\Omm{{\Omega_m}}

\def\Oml{{\Omega_{\Lambda}}}

\def\beq{\begin{equation}}
\def\eeq{\end{equation}}

\shorttitle{Quasar formation at $z \sim 6$}
\shortauthors{Li et al.}

\slugcomment{Accepted to ApJ}

\begin{document}

\title{Formation of $z \sim 6$ quasars
  from hierarchical galaxy mergers} 

\author
{
Yuexing Li\altaffilmark{1},
Lars Hernquist\altaffilmark{1}, 
Brant Robertson\altaffilmark{1,2,7},
Thomas J. Cox\altaffilmark{1}, 
Philip F. Hopkins\altaffilmark{1}, 
Volker Springel\altaffilmark{3}, 
Liang Gao\altaffilmark{4},
Tiziana Di Matteo\altaffilmark{5}, 
Andrew R. Zentner\altaffilmark{2},
Adrian Jenkins\altaffilmark{4},
Naoki Yoshida\altaffilmark{6}
}

\affil{$^{1}$Harvard-Smithsonian Center for Astrophysics, Harvard
  University, 60 Garden Street, Cambridge, MA 02138, USA}
\affil{$^{2}$Kavli Inst. for Cosmological Physics, Dept. of Astronomy
    \& Astrophysics, Univ. of Chicago, 933 East 56th Street, Chicago, IL
    60637, USA}
\affil{$^{3}$Max-Planck-Institute for Astrophysics,
  Karl-Schwarzschild-Str. 1, 85740 Garching, Germany} 
\affil{$^{4}$Inst. for Computational Cosmology, Dep. of Physics,
  Univ. of Durham, South Road, Durham  DH1 3LE, UK}
\affil{$^{5}$Dept. of Physics, Carnegie-Mellon University, 5000 Forbes
  Ave., Pittsburgh, PA 15213, USA} 

\affil{$^{6}$Nagoya University, Dept. of Physics, Nagoya, Aichi
  464-8602, Japan} 
\affil{$^{7}$ Spitzer Fellow}
\email{yxli@cfa.harvard.edu.}


\begin{abstract}

The discovery of luminous quasars at redshift $z \sim 6$ indicates the
presence of supermassive black holes (SMBHs) of mass $\sim 10^9\, \Msun$
when the Universe was less than one billion years old. This finding
presents several challenges for theoretical models, because whether
such massive objects can form so early in the $\Lambda$-cold dark
matter ($\Lambda$CDM) cosmology, the leading theory for cosmic
structure formation, is an open question.  Furthermore, whether the
formation process requires exotic physics such as super-Eddington
accretion remains undecided. Here, we present the first multi-scale
simulations that, together with a self-regulated model for the SMBH
growth, produce a luminous quasar at $z \sim 6.5$ in the $\Lambda$CDM
paradigm. We follow the hierarchical assembly history of the most
massive halo in a $\sim 3\, \Gpc^{3}$ volume, and find that this halo
of $\sim 8\times 10^{12}\, \Msun$ forming at $z \sim 6.5$ after several
major mergers is able to reproduce a number of observed properties of
SDSS J1148+5251, the most distant quasar detected at $z =6.42$
\citep{Fan2003}. Moreover, the SMBHs grow through gas accretion below
the Eddington limit in a self-regulated manner owing to feedback. We
find that the progenitors experience vigorous star formation (up to
$10^4\, \Msun\, \yr^{-1}$) preceding the major quasar phase such that
the stellar mass of the quasar host reaches $10^{12} \, \Msun$ at $z
\sim 6.5$, consistent with observations of significant metal
enrichment in SDSS J1148+5251.  The merger remnant thus obeys similar
$\mbulge$ scaling relation observed locally as a consequence of coeval
growth and evolution of the SMBH and its host galaxy. Our results
provide a viable formation mechanism for $z \sim 6$ quasars in the
standard $\Lambda$CDM cosmology, and demonstrate a common,
merger-driven origin for the rarest quasars and the fundamental
$\mbulge$ correlation in a hierarchical Universe.

\end{abstract}

\section{INTRODUCTION}

Quasars rank among the most luminous objects in the Universe and are
believed to be powered by SMBHs (e.g., \citealt{Salpeter1964,
Lynden-Bell1969}). They constrain the formation and evolution of
galaxies and SMBHs throughout cosmic time. The similarity between the
cosmic star formation history (e.g., \citealt{Madau1996, Bunker2004,
Bouwens2004}) and the evolution of quasar abundances (e.g.,
\citealt{Shaver1996}) suggests an intriguing link between galaxy
formation and black hole growth. This is strengthened by tight
correlations measured locally between the masses of the black holes
and the global properties of the spheroid components of their hosts,
such as their luminosities and masses (\citealt{Magorrian1998,
Marconi2003}), light concentration (\citealt{Graham2001}), and
velocity dispersions (\citealt{Ferrarese2000, Gebhardt2000,
Tremaine2002}).

Distant, highly luminous quasars are important cosmological probes for
studying the first generation of galaxies, the star formation history and
metal enrichment in the early Universe, the growth of the first
supermassive black holes (SMBHs), the role of feedback from quasars
and black holes in galaxy evolution, and the epoch of reionization. The Sloan
Digital Sky Survey (SDSS, \citealt{York2000}) has contributed significantly to
the discovery of high redshift quasars.  Currently, there are over 1000
quasars known at $z \gtrsim 4$ and 12 at $z \gtrsim 6$ (\citealt{Fan2001,
  Fan2003, Fan2004, Fan2006A}). As reviewed by \cite{Fan2006B}, quasars at $z
\sim 6$ are characterized by: (1) a low space density ($\sim 10^{-9}\,
\Mpc^{-3}$ comoving); (2) high luminosities (absolute luminosity at 
rest-frame $M_{1450\AA} < -26$), believed to be powered by SMBHs of
$\sim 10^9\, \Msun$; (3) Gunn-Peterson absorption troughs
\citep{Gunn1965} in their spectra, which place these quasars at the
end of the epoch of reionization (e.g., \citealt{Fan2001, Becker2001,
Djorgovski2001, Lidz2002, Songaila2002, White2003, Sokasian2003}); and
(4) a lack of evolution in the spectral energy distribution compared
to lower-shift counterparts (e.g., \citealt{Elvis1994, Glikman2006,
  Richards2006}), which demonstrates the existence of ``mature'' quasars at
early times and comparable metal enrichment in quasars at all cosmic epochs.

The most distant quasar known, SDSS J1148+5251 (hereafter
J1148+5251), was discovered by SDSS at z = 6.42 \citep{Fan2003}. It is
extremely bright optically with $M_{1450\AA}$ = -27.8, and deep imaging
surveys in both optical and radio (\citealt{Carilli2004, White2005,
  Willott2005}) show no sign of gravitational lensing or other companions at
the same redshift in the vicinity. Over the past few years, this quasar has
been extensively studied at many wavelengths. Near-infrared observations by
\cite{Willott2003} and \cite{Barth2003} imply a bolometric luminosity of
$\Lbol \sim 10^{14}\, \Lsun$ powered by accretion onto a SMBH of mass 1--5
$\times 10^9 \, \Msun$. Radio observations by \cite{Bertoldi2003A} and
\cite{Carilli2004} suggest that the host is a hyper-luminous far-infrared
(FIR) galaxy, with $\Lfir \sim 10^{13}\, \Lsun$, and these authors estimate a
star formation rate of $\sim 3\times 10^3\, \Msun\, \yr^{-1}$ by assuming most
of the FIR luminosity comes from young stars.  Emission from carbon monoxide
(CO) has been detected (\citealt{Walter2003, Bertoldi2003B, Walter2004}) 
corresponding to a mass of $\sim 2\times 10^{10} \, \Msun$.  Dust has been
seen by several groups (e.g., \citealt{Robson2004, Bertoldi2003A, Carilli2004,
  Beelen2006}) with an estimated mass of $\sim 5\times 10^8\, \Msun$.  In
particular, {\em Spitzer} observations by \cite{Charmandaris2004} and
\cite{Hines2006} indicate that the dust is heated by an active galactic
nucleus (AGN). Furthermore, the detection of iron by \cite{Barth2003}, the
carbon fine structure line [CII] by \cite{Maiolino2005} and excess OI
absorption by \cite{Becker2006} indicate near-solar metallicity in the quasar
host. 

These observations raise many fundamental questions for models of 
quasar and galaxy formation: Where do such high-redshift, luminous quasars
originate? How do they form? What are the mechanisms and physical conditions
for SMBH growth? And, do these quasar hosts obey the same SMBH--host
correlations as observed in the local Universe?  

Interpretations of various observations of \zquasar\ have painted a rather
conflicting picture for the formation site of the quasar halo and the
SMBH--host relationship. The low abundance of these quasars leads to the
view that they are hosted by massive halos ($\gtrsim 10^{13}\, \Msun$) in the 
rarest density peaks of the dark matter distribution \citep{Fan2003}. However,
it has been argued, based on the lack of companion galaxies in the field, that
this quasar may reside in a much lower mass halo in a less overdense region
(\citealt{Carilli2004, Willott2005}). Moreover, \cite{Walter2004} suggest,
based on the dynamical mass estimate from CO measurements, that \zquasar\
contains a small stellar spheroid, and that the SMBH may have largely formed
before the host galaxy. However, the detection of metal lines
(\citealt{Walter2004, Barth2003, Maiolino2005}), along with dust
(\citealt{Bertoldi2003A, Carilli2004, Robson2004, Charmandaris2004, Hines2006,
Beelen2006}), indicates that the interstellar medium (ISM) of \zquasar\ was 
significantly enriched by heavy elements produced through massive star
formation at rates of $\sim 10^3\, \Msun\, \yr^{-1}$ occurring as early as $z
\gtrsim 10$, and that large stellar bulges form before accreting SMBHs
undergo luminous quasar activity.   

In an expanding Universe that is dominated by cold dark matter and is
accelerated by dark energy, the $\Lambda$-cold dark matter ($\Lambda$CDM)
cosmology, the leading theoretical model for structure formation, assumes
that structure grows from weak density fluctuations amplified by gravity, with
small objects collapsing first and subsequently merging to form progressively
more massive ones, a process known as ``hierarchical assembly'' (e.g., see
\citealt{Barkana2001} for a review). The formation of galaxies and quasars is 
therefore determined by the abundance of dark matter halos; i.e., the
number density of halos as a function of mass and redshift. The most widely
used analytic model for the halo mass function was first developed by
\cite{Press1974} (hereafter PS), which is based on Gaussian density 
perturbations, linear gravitational growth, and spherical collapse of dark  
matter. Using the PS mass functions, \cite{Efstathiou1988} studied the 
abundance of rare objects, such as luminous quasars at high redshifts. 
These authors predicted a sharp ``cut-off'' of the quasar population at $z
\sim 5$. However, while the initial, linear growth of density perturbations
can be calculated analytically, the gravitational collapse and subsequent
hierarchical build-up of structure is a highly nonlinear process that can be
followed only through numerical simulations. It has been shown by
previous numerical studies (e.g., \citealt{Jenkins2001, Sheth2002,
  Springel2003b}), and more recently by the state-of-the-art {\em Millennium
  Simulation} by \cite{Springel2005A}, that the PS formula underestimates the
abundance of high-mass halos by up to an order of magnitude. Therefore,
whether or not rare quasars such as \zquasar\ can form in the $\Lambda$CDM
cosmology remains an open question and an important test of the theory.  

To date, only a limited number of analytical or semi-analytic models have
addressed the early formation of a $10^9 \, \Msun$ SMBH at $z \sim 6$  
(\citealt{Haiman2001, Haiman2004, Yoo2004, Volonteri2005}). These approaches
use merger trees of dark matter halos generated using the PS theory, and
assume a black hole accretion rate at or above the Eddington limit. However,
as discussed above, the PS--based approach may be inaccurate. Moreover,
it is not clear whether sufficient physical conditions for such large
accretion rates exist in quasar systems as the hydrodynamics of the 
large-scale gas flow and feedback from black holes have not been incorporated    
in earlier modeling.

It is believed that the growth of SMBHs is closely linked
to galaxy formation (e.g., \citealt{Magorrian1998, Ferrarese2000,
Gebhardt2000, Graham2001, Tremaine2002, Marconi2003, Haiman2004,
Kazantzidis2005, Li2006A}), and that the growth is self-regulated 
by feedback (e.g., \citealt{Silk1998, Haehnelt1998A, Fabian1999, King2003,
Wyithe2003, DiMatteo2005, Springel2005B, Sazonov2005, Murray2005,
Wyithe2005}). Remarkably, self-regulated models with SMBH feedback in the form
of thermal energy coupled to the ambient gas have been demonstrated to
successfully reproduce many observations of galaxies, including the $\msigma$
relation (\citealt{DiMatteo2005, Robertson2006A}), galaxy colors
(\citealt{Springel2005C, Hopkins2006F}), X-ray gas emission \citep{Cox2006a},
elliptical kinematics \citep{Cox2006b} and the fundamental plane
\citep{Robertson2006B}, quasar properties (\citealt{Hopkins2005A,
  Hopkins2005B}), luminosity functions (\citealt{Hopkins2005C, Hopkins2005D,
  Hopkins2006B}), and populations (\citealt{Hopkins2006A, Hopkins2006C,
  Hopkins2006D}), and the luminosity function of low-level AGN
\citep{Hopkins2006E}. Furthermore, these simulations of binary mergers
identify a plausible, merger-driven formation mechanism for massive black
holes and luminous quasars (e.g., \citealt{DiMatteo2005, Hopkins2006A,
  Robertson2006A}).  

Here, we present a model that accounts for the SMBH growth, quasar activity
and host galaxy properties of the most distant quasar known. In our scenario,
the quasar and its host galaxy form in a  massive halo that originates from a
rare density peak in the standard $\Lambda$CDM paradigm, and they grow
hierarchically through multiple gas-rich mergers, supporting an average star
formation rate of $\sim 10^3\, \Msun\, \yr^{-1}$ that peaks at $z\sim
8.5$. Once the progenitors undergo sufficient dynamical friction to 
coalesce, the multiple SMBHs from the progenitor galaxies merge and
exponentially increase their mass and feedback energy via accretion. At $z
\approx 6.5$ when the SMBH mass exceeds $10^{9}\, \Msun$, black hole accretion
drives a sufficiently energetic wind to clear obscuring material from the
central regions of the system and powers an optically luminous quasar similar
to \zquasar. We have devised a set of novel multi-scale simulations, which
include cosmological N-body calculations on large scales and hydrodynamic
simulations of galaxy mergers on galactic scales, coupled with the
self-regulated growth of SMBHs, enabling us to follow galaxy assembly and
quasar formation at $z \sim 6$.     

This paper is organized as follows. In \S~2, we describe
our computational methods and models, which includes a set of large scale
cosmological N-body simulations, and hydrodynamical galaxy mergers along the
merging history of the quasar halo. In \S~3, we present the formation and
evolution of the quasar and its host galaxy, including the assembly of the
galaxy progenitors, star formation, and SMBH growth, as well
as the SMBH--host correlations, and properties of the quasar such as
luminosities and lifetimes. We discuss feedback from starburst-driven winds, 
quasar abundances for cosmological models with different parameters, the
implication of black hole mergers, and galaxies in the epoch of reionization
in \S~4, and summarize in \S~5.    

\section{Methodology}

Rare, high-redshift quasars originate in highly overdense regions in the
initial dark matter density distribution and grow through hierarchical
mergers, as predicted by the $\Lambda$CDM theory. Simulations of high-redshift
quasar formation must model a large cosmological volume to accommodate the low
abundance of this population, have a large dynamic range to follow the
hierarchical build-up of the quasar hosts, and include the hydrodynamics of
the gas flows in galaxy mergers. The cutting-edge {\em Millennium Simulation}
by \cite{Springel2005A} follows structure formation in a box with side length
of $500\, h^{-1}\, \Mpc$ using $2160^3$ dark matter particles. It reproduces
the large-scale galaxy distribution as observed \citep{Springel2006}, and
identifies an early quasar halo candidate at $z = 6.2$ which ends up in the
richest cluster at the present day. However, even such a large dynamic range
still falls short of being able to follow the formation and evolution of the
rarest quasars observed at the highest redshifts. Moreover, in order to
address the properties of quasars and host galaxies, gasdynamics and physical
processes related to star formation and black hole growth must be included.
To satisfy these requirements, we have performed a set of novel multi-scale
simulations that enable us to resolve individual mergers on galactic scales
and retain the context of large-scale structure formation, as well as the
evolution of black holes and stars.     

First, we perform a coarse dark matter simulation in a volume
of $1\, h^{-3}\, \Gpc^{3}$ designed to accommodate the low number density
of $z\approx6$ quasars. The largest halo at $z=0$, within which the
descendants of early, luminous quasars are assumed to reside
\citep{Springel2005A}, is then selected for resimulation with higher
resolution using a multi-grid zoom-in technique developed by
\cite{Gao2005}. The merging history of the largest halo 
at $z \sim 6$, which has reached a mass of $\sim 5.4 \times10^{12}\, h^{-1}\, 
\Msun$ through 7 major (mass ratio $<$ 5:1) mergers between redshifts 14.4 and
6.5, is extracted. These major mergers are again re-simulated hydrodynamically
using galaxy models scaled appropriately for redshift \citep{Robertson2006A} and
adjusted to account for mass accretion through minor mergers. The simulations
include prescriptions for star formation \citep{Springel2003a}, and for SMBH
growth and feedback (\citealt{DiMatteo2005, Springel2005B}), as
described below. 

\subsection{Code and Parameters}

\begin{figure*}
\begin{center}
\includegraphics[width=7in]{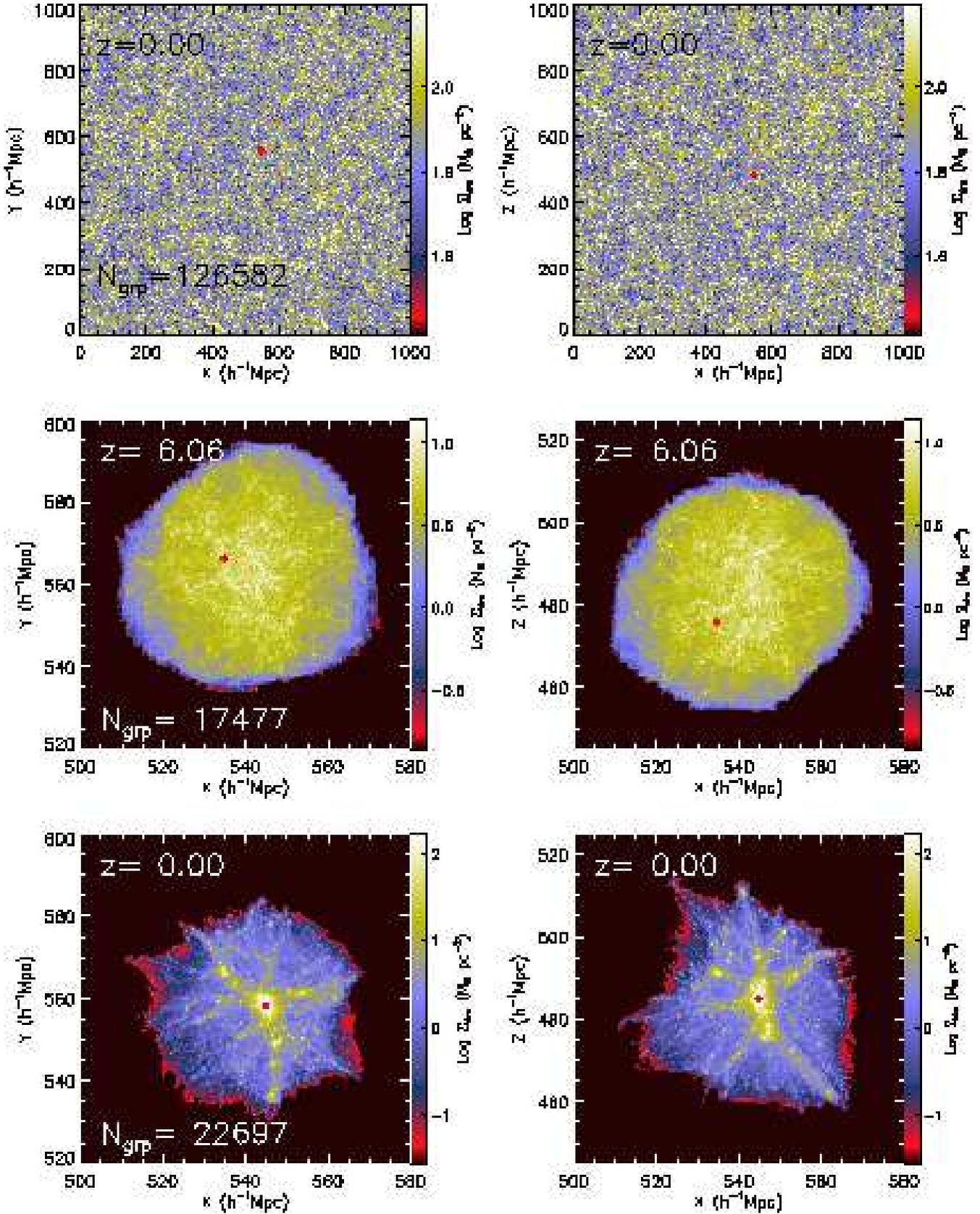}
\vspace{0.5cm}
\caption{Snapshots from a cosmological simulation run with WMAP1
  parameters. The images show projected density of dark matter in x-y
  (\textit{left column}) and x-z (\textit{right column}) planes, the red dot
  represents the center-of-mass of the quasar halo, which is the largest halo
  at both $z=0$ and $z=6$. The top panels show the coarse run at $z=0$. The
  middle and bottom  panels show the zoom-in run at $z=6.06$ and $z=0$,
  respectively, the number at the lower-left corner indicates the number of
  groups identified at that redshift.}  
\label{FigMap}
\end{center}
\end{figure*}

Our multi-scale simulations were performed using the parallel, N-body/Smoothed
Particle Hydrodynamics (SPH) code GADGET2 developed by \cite{Springel2005D}
that is well tested in a wide range of applications from large scale structure
formation to star formation. For the computation of gravitational forces, the
code uses the ``TreePM'' method \citep{Xu1995} that combines a ``tree''
algorithm \citep{Barnes1986} for short-range forces and a Fourier transform 
particle-mesh method \citep{Hockney1981} for long-range forces. The PM-method 
works efficiently in large scale cosmological simulations, while 
the tree-method provides accurate forces for the large dynamic range
of galaxy merger simulations.

GADGET2 implements the entropy-conserving formulation of SPH
\citep{Springel2002} with adaptive particle smoothing, as in
\cite{Hernquist1989A}. Radiative cooling and
heating processes are calculated assuming collisional ionization equilibrium
\citep{Katz1996, Dave1999}. Star formation is modeled in a multi-phase ISM,
with a rate that follows the Schmidt-Kennicutt Law (\citealt{Schmidt1959,
  Kennicutt1998}). Feedback from supernovae is captured through a multi-phase
model of the ISM by an effective equation of state for star-forming gas
\citep{Springel2003a}. A prescription for supermassive black hole growth and 
feedback is also included, where black holes are represented by collisionless
``sink'' particles that interact gravitationally with other components and
accrete gas from their surroundings. The accretion rate is estimated from the
local gas density and sound speed using a spherical Bondi-Hoyle
\citep{BondiHoyle1944, Bondi1952} model that is limited by the Eddington rate.
Feedback from black hole accretion is modeled as thermal energy injected into
surrounding gas, as described in \cite{Springel2005B} and \cite{DiMatteo2005}.    

The simulations presented in this paper adopt the $\Lambda$CDM model with
cosmological parameters chosen according to the first year Wilkinson Microwave
Anisotropy Probe data (WMAP1, \citealt{Spergel2003}), ($\Omega_{\rm{m}}$,
$\Omega_{\rm{b}}$, $\Omega_{\Lambda}$, $h$, $n_s$, $\sigma_8$)= (0.3, 0.04,
0.7, 0.7, 1, 0.9). Here,
$\Omega_{\rm{m}}$ is the total matter density in units of the
critical density for closure, $\rho_{\rm{crit}} =3 H_0^2/(8\pi
G)$. Similarly, $\Omega_{\rm{b}}$ and $\Omega_\Lambda$ denote the
densities of baryons and dark energy at the present day. The Hubble 
constant is parameterized as $H_0 = 100\, h\, {\rm{km}\, \rm s^{-1}\,
  \Mpc^{-1}}$, while $\sigma_8$ is the root-mean-squared linear mass 
fluctuation within a sphere of radius $8\, h^{-1}\, \Mpc$ extrapolated
to $z=0$. We have also done the same cosmological simulations with WMAP third 
year results (WMAP3, \citealt{Spergel2006}), ($\Omega_{\rm{m}}$,
$\Omega_{\rm{b}}$, $\Omega_{\Lambda}$, $h$, $n_s$, $\sigma_8$)= (0.236, 0.042,
0.759, 0.732, 0.95, 0.74) for comparison.    

\subsection{Cosmological Simulations}

The quasars at $z \sim 6$ have an extremely low comoving space
density, $n \sim 10^{-9}\, \Mpc^{-3}$, and are believed to
reside in massive dark matter halos with $M\gtrsim 10^{13}\, \Msun$
\citep{Fan2003}.  Cosmological simulations of quasar 
formation must therefore model a volume of $\sim1\, h^{-3}\, \Gpc^{3}$ to
account for the rarity of such objects. However, in 
order to resolve a $10^{13}\, \Msun$ halo at $z \sim 6$ in a
cosmological simulation with uniform resolution, a dark matter particle
mass at least as small as $10^{11}\, h^{-1}\, \Msun$ and particle numbers of $>
10^9$ are required. Tracking the merger history of such halos requires
$\sim 2$ orders of magnitude higher resolution and would be
computationally prohibitive with standard techniques.

We achieve the mass resolution requirements for the merger history of a
$10^{13}\, \Msun$ halo at $z\sim 6$ by means of a two-step re-simulation.  
First, coarse dark matter cosmological simulations are performed to
identify a candidate halo for the quasar host. A cubic volume
$L_{\rm{box}} = 1\, h^{-1}\, \Gpc$ on a side is simulated with
$400^3$ particles, achieving mass and force resolutions of $\mdm \sim
1.3\times 10^{12}\, h^{-1}\, \Msun$ and $\epsilon \sim 125\, h^{-1}\, \kpc$
(comoving), respectively. To generate the initial conditions, we 
use the Boltzmann code {\small CMBFAST} by \cite{Seljak1996} to compute a
linear theory power spectrum for our chosen cosmology. A random realization of
the power spectrum is constructed in Fourier space, sampling modes in a sphere
up to the Nyquist frequency of the mesh. The particle distribution is evolved
forward in time to $z=0$ from its initial displacement at $z=30$ determined 
using the Zel'dovich approximation.

At the end of the simulation, halos are identified using a ``friends-of-friends''
(FOF) group-finding algorithm \citep{Davis1985} with a fixed comoving linking
length equal to 0.2 times the mean dark matter interparticle separation and a
minimum of 32 particles per group \citep{Springel2003b}. The mean overdensity
of the groups corresponds approximately to the expected density of virialized
halos \citep{Springel2005A}. From the more than 126000 groups identified in the $1\,
h^{-3} \, \Gpc^{3}$ volume at $z=0$ the largest halo with $M(z=0)
\simeq 3.6 \times 10^{15}\, h^{-1}\, \Msun$ is selected as a candidate halo
for modeling the formation of a quasar at $z=6.5$.

A multi-grid technique developed by \cite{Gao2005} and \cite{Power2003} is
used to `` zoom in'' with high resolution 
on the selected halo region which has an effective side length of
$L_{\rm{box}} \sim 50\, h^{-1}\, \Mpc$. Large-scale tidal forces are captured
by binning exterior particles into cells according to their distance from the
high-resolution region. To ensure proper treatment of small-scale structure,
the initial displacements of the high resolution particles are calculated
assuming a higher initial redshift of $z=69$ and normalized to $\sigma_8$ at
$z=0$. The re-simulation uses $\approx350^3$ particles, with $\approx 340^3$
particles inside the high-resolution region. With this technique, the mass
resolution increases by almost four orders of magnitude to $\mdm \sim
2.8\times 10^{8}\, h^{-1}\, \Msun$ while the spatial resolution reaches
$\epsilon \sim 5\, h^{-1}\, \kpc$. 

Figure~\ref{FigMap} shows snapshots of both the coarse and high-resolution
zoom-in runs that locate the quasar halo candidate. In the coarse run, the
``cosmic web'' is clearly seen, although the distribution appears nearly
homogeneous on such large scales. In the zoom-in run, filamentary structures are
prominent. Dark matter collapses along the filaments, and the largest halo
forms in the deepest potential wells at the intersections of the filaments. 
The high resolution of the zoom-in run enables the identification of more halos
with lower masses both at $z=0$ and at high redshifts as early as $z \sim
17$, which is sufficient to identify the halo progenitors of the candidate
quasar at $z \sim 6$. It appears that the halo progenitor of the largest one
at the present day is also the most massive halo at $z \sim 6$, when it reaches a
mass of $M\approx 5.4 \times 10^{12}\, h^{-1}\, \Msun$, making it a plausible
candidate for hosting a rare $z \sim 6.4$ quasar.   

\subsection{Halo Mass Functions with Different Cosmological Parameters}
\label{subsec_mf}

\begin{figure*}
\hspace{-1cm} \includegraphics[width=7in]{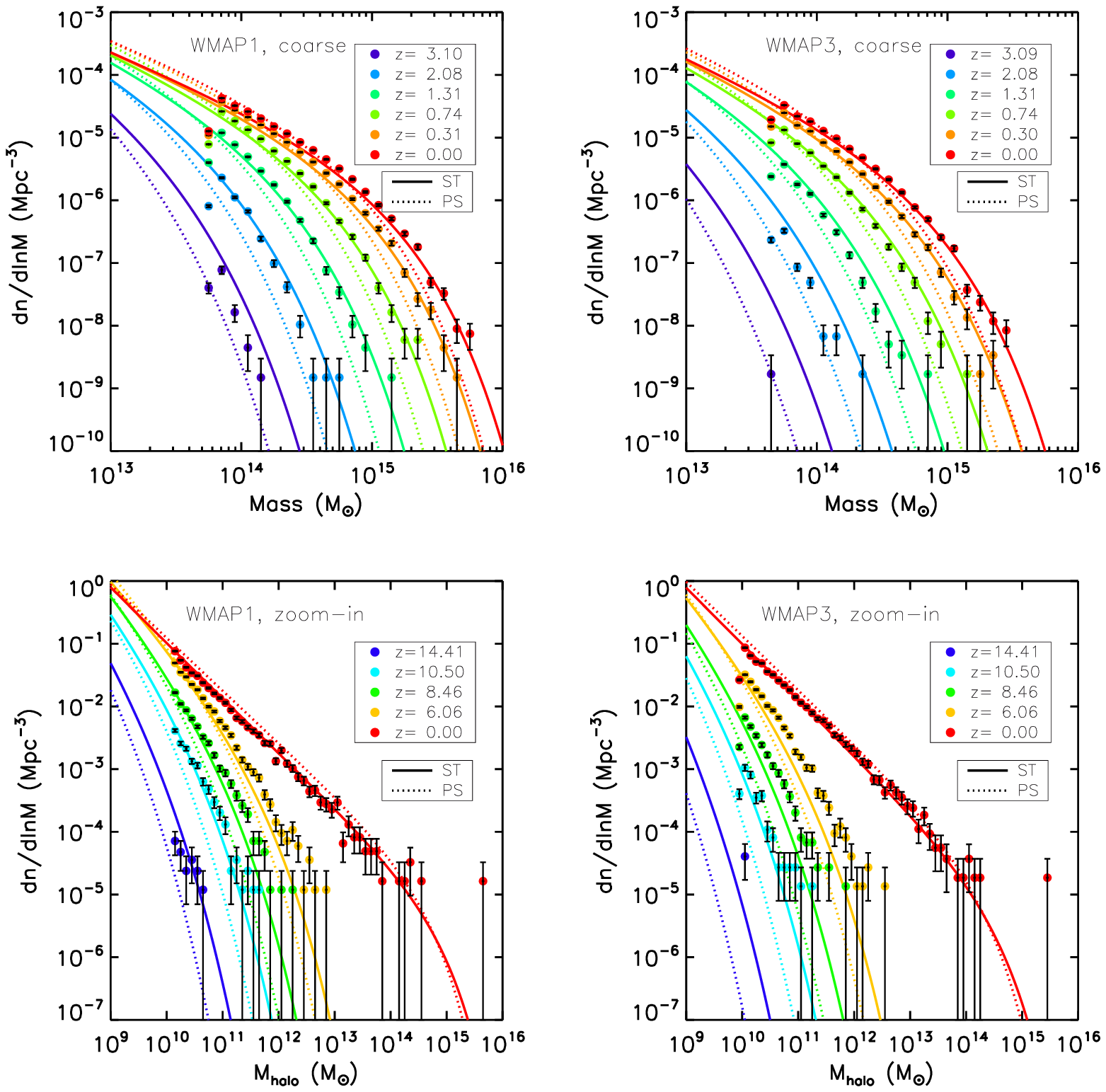}
\vspace{-5cm}
\caption{Halo mass functions from cosmological simulations with
  parameters from WMAP1 ({\em left}) and WMAP3 ({\em right}),
  respectively, and with different levels of resolution corresponding
  to our coarse ({\em top panels}) and zoom-in ({\em bottom panels}) runs. The 
  colored symbols indicate different redshifts, while the error bar shows
  the Poisson error $\sqrt{N}$. The mass
  functions from \cite{Press1974} (PS, dotted line) and  \cite{Sheth2002} (ST,
  solid line) are also shown for comparison. Please note that in the bottom
  panels, the analytical curves (ST and PS) apply only to a {\em random}
  region, they are not suitable for a highly overdense region where the most
  massive halos reside in the zoom-in box. So the high-mass end of the
  simulated mass function deviates significantly from the prediction,
  see text for more discussion.} 
\label{FigMF}
\end{figure*}

The impact that variations in the cosmological parameters can have on
large-scale structure formation can be understood from the theoretical mass
function of halos, as derived by \cite{Press1974} and later developed by
\cite{Lacey1993}. The comoving number density ${\rm d} n$ of halos of mass
between $M$ and $M+ {\rm d} M$ can be described as,   
\begin{equation}
\frac{{\rm d}\, n}{{\rm d}\, M}= \sqrt{\frac{2}{\pi}}\, \frac{\rho_0}{M^2}\,
\frac{\delta_{\rm c}(z)}{\sigma(M)} \left|\frac{\rm{d}\,\ln \sigma}{{\rm d}\,
  \ln M}\right| \exp\left[-\frac{\delta_{\rm c}(z)^2}{2\, \sigma^2(M)}\right] \, ,
\label{eq_mf}
\end{equation}
where $\rho_0$ is the local mean mass density, $\delta_{\rm c}(z)$ is the
critical density of collapse at redshift $z$ linearly extrapolated to the
present day, and $\sigma(M)$ is the mass variance, which is a function of
the power spectrum $P(k)$ with wavenumber k and the window
function $w(k)$, $\sigma^2(M)= \frac{1}{2 \pi^2} \int_0^{\infty}\, P(k)\,
w^2(k)\, d^3k$. The abundance of halos depends on the two functions
$\sigma(M)$ and $\delta_{\rm crit} (z)$, each of which involves the
cosmological parameters, in particular $\sigma_8$, $\Omm$ and $\Oml$. These
parameters determine the formation epoch of a halo and its mass.  

The recently-released third year WMAP3 results \citep{Spergel2006}
have lower values of $\sigma_8$, $n_{\rm s}$ and $\Omm$, compared to WMAP1
\citep{Spergel2003}. The smaller $\sigma_8$ from WMAP3 would lower the
amplitude of the power spectrum, which in turn reduces
$\sigma(M)$. Furthermore, a smaller $\Omm$ would reduce $\delta_{\rm c}(z)$ 
and hence delay halo formation. So, compared to WMAP1, at a given redshift the
WMAP3 parameters would yield a lower abundance of halos 
with mass ${M_{\rm {halo}} \gtrsim M_{*}}$, where ${M_{*}}$
is the halo mass corresponding to the characteristic luminosity in the
Schechter luminosity function for galaxies \citep{Schechter1976}; while
for $M_{\rm {halo}} < M_{*}$, it predicts a slightly larger halo abundance.  
  
To test the sensitivity of our model to the new WMAP results, we have
performed the same set of cosmological simulations with parameters
from the WMAP3 measurements \citep{Spergel2006}. We find that indeed
the changes implied by the new parameters primarily affect the
formation time and the mass of the candidate quasar halo. For the same
random phases in the initial conditions, the location of the most
massive halo at $z=0$ remains the same in both the WMAP1 and WMAP3
runs, except that its mass is reduced by a factor of $\sim 1.6$ for the
WMAP3 parameters. Similarly, the mass of the largest halo at $z
\approx 6$ is altered by roughly the same factor.  Other notable
changes include: (1) the formation epoch of the first halo is shifted
from $z \sim 16.8$ in the WMAP1 run to $z \sim 14.4$ in the WMAP3 run,
and (2) the merging history of the largest halo at $z \sim 6$ moves to
lower redshifts in the WMAP3 run, but the number of major mergers
remains the same.

Figure~\ref{FigMF} shows the halo mass functions from different
cosmological simulations. The PS mass function \citep{Press1974}, as
well as the one corrected to match numerical simulations by
\cite{Sheth2002} (ST), are also shown for comparison. One important
feature in this figure is that the coarse runs agree well with the ST
mass function, but show a larger comoving density at the high mass end
than that predicted by the PS theory. Our results show that the PS
formula underestimates the abundance of high-mass halos by nearly an
order of magnitude at $z =0$, and the discrepancy between the PS
calculation and numerical simulations becomes larger at higher
redshifts, confirming previous findings (e.g., \citealt{Jenkins2001,
Sheth2002, Springel2003b, Springel2005A}). This may explain why previous
models using the PS formula to study the abundance of luminous quasars, which    
presumably form in massive halos, under-predicted the number of bright
quasars at $z>5$ (e.g., \citealt{Efstathiou1988}). Furthermore, these
results also suggest that the commonly used analytical merger tree
generated using the PS formula may not be suitable to study quasar
formation at high redshifts.

There are two clear ``shifts'' of the mass function caused by
resolution and cosmological parameters. Those from runs with higher
resolution extend to higher redshifts, and at the same redshift, the
WMAP1 runs produce more massive halos than the WMAP3 ones. As shown in
Figure~\ref{FigMF}, the coarse runs produce mass functions only up to
$z \approx 3$ owing to limited mass resolution, while the zoom-in runs
can produce quite reasonable mass functions as early as $z \approx
14$. Because the zoom-in runs were deliberately centered on the
highest density peak of the $1\, h^{-3}\, \Gpc^{3}$ box, they each
contain a very massive halo ($M > 10^{15}\, \Msun$ at $z=0$) by
construction. This explains why the highest mass bin (which contains only one  
halo in this case) is  $\sim 2$ orders of magnitude larger than the
theoretical curves (ST, PS), which apply only to a {\em random} region that
has a much lower density fluctuation.  

To summarize, at a given redshift, runs with the WMAP3 parameters
yield slightly less massive halos than ones performed with the WMAP1
values.  Or, to put it differently, objects in the WMAP3 cosmology
will have masses similar to those for WMAP1, but at slightly later
times (i.e. lower redshifts).  In what follows, we are primarily
concerned with investigating the plausibility of forming $z\sim 6$
quasars through the self-regulated growth of SMBHs in hierarchical
mergers, rather than precisely reproducing the properties of an
individual quasar at a given redshift, such as J1148+5251. Most of our 
results are therefore based on runs with the WMAP1 parameters, to ease
comparison with previous numerical work. If it were firmly established that
e.g. $\sigma_8$ is in reality smaller than its WMAP1 value, then a more exact
match to a particular quasar could presumably be obtained by considering
larger simulation volumes and identifying a suitable candidate host that is
slightly rarer than the one we have chosen to focus on here.

\subsection{Merger Tree Construction}

\begin{figure}
\begin{center}
\includegraphics[width=3.5in]{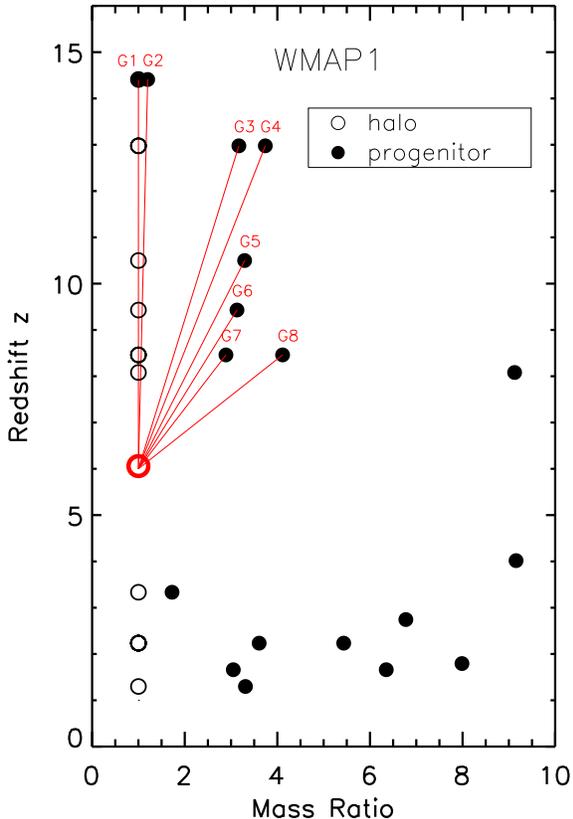}
\caption{
Schematic merging history of the largest halo at $z=0$ traced by 
mergers at different redshifts with mass ratio $\le 10:1$, which is defined
as the mass ratio between the halo and progenitor at a given time. Each of
the progenitors joins in this big merging event at a given redshift,
interacts with the system, and subsequently merges with others at later
times. The quasar host at $z \sim 6$ is built up by seven successive major
mergers of progenitors G1, G2, ... G8 from $z \sim 14.4$ to $z \sim 6.0$,
as illustrated by the red lines in this plot. The first interaction between G1
and G2 takes place at $z \sim 14.4$, then G3 and G4 join in the system at $z
\sim 13$, followed by G5, G6, G7 and G8 at later times (see text for more
details). The timeline of these events, the mass and other properties of
these progenitors are listed in Table~1.}
\label{FigTree}
\end{center}
\end{figure}

To follow the hierarchical mass assembly of the host galaxy over cosmic time,
the merger tree of the halo is extracted from the cosmological
simulation. This tree provides key information for computing the physical
properties of the progenitor galaxy population. While the merger history of
the halo includes a spectrum of progenitor masses, the most massive
progenitors contribute the majority of the halo mass over the redshift range
considered. We trace the merger history of the most massive progenitor at each
redshift by using particle tags to identify progenitor systems at earlier
redshifts in the simulation. Groups that contribute at least 10\% of the halo
mass at a given time step are considered as the progenitors of the halo and
are recorded. The procedure is repeated until the last progenitor is reached, 
producing the merging history.

Figure~\ref{FigTree} illustrates the merging history of the largest halo at
$z=0$ in our cosmological simulation, which has a mass of $\simeq 3.6
\times 10^{15}\, h^{-1}\, \Msun$. It is also the largest one at $z\sim6$ with
a mass of $\simeq 5.4 \times 10^{12}\, h^{-1}\, \Msun$. This schematic plot
outlines the redshift of merger event, and the mass ratio of the halo to its
galaxy progenitors at a given time. It shows that this halo grows rapidly
through hierarchical mergers early on, with seven major mergers (mass ratio of
the merging pairs $\le 5:1$) from $z \sim 14.4$ to $z \sim 8.5$ that build up a
substantial fraction of the halo mass at $z \sim 6$. 

In modeling the development of a $z\sim 6$ quasar, we are primarily
interested in ``major'' mergers, where the mass ratio of the merging
galaxies is not too far from unity, for several reasons.  First, it is
believed that major mergers play the most important role in the
formation and evolution of massive galaxies (e.g., \citealt{Sanders1996,
  Scoville2000, Veilleux2002, Conselice2003, Dasyra2006}). Second, and of
greater concern to us in this paper, in our picture for quasar fueling, gas in
a rotationally supported disk loses angular momentum through gravitational
torques excited by tidal forces in a merger, driving the growth of
supermassive black holes. This process operates most effectively in a major
merger because the tidal deformation of each galaxy is significant in such an
event \citep{Barnes1991, Barnes1992, Barnes1996}.  Collisions involving galaxies
with a mass ratio as large as $10:1$ can induce gas inflows in disks
\citep{Hernquist1989B, Hernquist1995}, but only for limited orbital
configurations.  For these reasons, we focus on mergers from the merger tree
having a mass ratio $\le 5:1$, as outlined by the red color in
Figure~\ref{FigTree}. 

In the resimulation of the merger tree as described in \S~\ref{sec_mt}, we
take into account mass accretion of the halo by adding mass proportionally to
each of the eight progenitor galaxies in the major mergers. This approach
preserves the progenitor mass ratios and approximately preserves the dynamics
of the major mergers \citep{Dubinski1998}. 

\subsection{Simulations of Galaxy Mergers Along the Tree}
\label{sec_mt}

\begin{deluxetable*}{lccccccc}
\label{tab1}
\tablecolumns{8}
\tablecaption{Progenitor Properties and Numerical Parameters}
\tablehead{
\colhead{Galaxy\tablenotemark{a}}    & 
\colhead{$z$\tablenotemark{b}}       &
\colhead{$\Mvir$\tablenotemark{c}}   &
\colhead{$\Vvir$\tablenotemark{d}}   &
\colhead{$f_{\rm{gas}}$\tablenotemark{e}}   & 
\colhead{$\MBH$\tablenotemark{f}}    &
\colhead{$R_{\rm p}$\tablenotemark{g}}  &
\colhead{$R_0$\tablenotemark{h}} \\
&
&
\colhead{[$10^{10}\, h^{-1}\, \Msun$]} &
\colhead{[km s$^{-1}$]}           &
&
\colhead{[$10^{5}\, h^{-1}\, \Msun$]}&
\colhead{[$h^{-1}\, \kpc$]}&
\colhead{[$h^{-1}\, \kpc$]}
}  
\startdata
 G1 & 14.4 &   6.3 & 234.1 & 1.0 & 0.15 & --  & --  \\
 G2 & 14.4 &   5.3 & 220.3 & 1.0 & 0.15 & 0.2 & 7.1  \\
 G3 & 13.0 &  15.0 & 297.8 & 1.0 & 0.51 & 0.2 & 8.5  \\
 G4 & 13.0 &  17.7 & 314.6 & 1.0 & 0.51 & 0.3 & 10.7 \\
 G5 & 10.5 &  49.1 & 401.0 & 1.0 & 6.56 & 0.4 & 11.3 \\
 G6 & 9.4  &  79.6 & 448.6 & 0.9 & 23.3 & 0.5 & 18.2 \\
 G7 & 8.5  & 160.0 & 540.4 & 0.9 & 89.2 & 0.7 & 25.2 \\
 G8 & 8.5  & 207.7 & 589.5 & 0.9 & 89.2 & 1.0 & 34.5 \\ 
\enddata 
\tablenotetext{a}{Name of galaxy progenitor. G1 is the halo at $z=14.4$.}
\tablenotetext{b}{Redshift at which the progenitor enters the merger tree.}
\tablenotetext{c}{Virial mass, assuming overdensity $\Delta=200$.}
\tablenotetext{d}{Virial velocity, assuming overdensity $\Delta=200$.}
\tablenotetext{e}{Gas fraction of the progenitor baryonic mass.}
\tablenotetext{f}{Progenitor black hole mass at the merger redshift.}
\tablenotetext{g}{Pericentric distance of the incoming progenitor to the
 center-of-mass of the existing system.}
\tablenotetext{h}{Initial distance of the incoming progenitor to the
  existing system.}
\end{deluxetable*}

In order to model the formation and evolution, and properties of the quasar
candidate, the merger tree is then re-simulated hydrodynamically with
galaxy models that consist of an extended dark matter halo, a rotationally
supported, exponential disk of gas and stars and a central supermassive black
hole. We follow the evolution of the system built up by seven major mergers
hierarchically from $z \sim 14.4$ to $z \sim 6$, as shown in
Figure~\ref{FigTree}. Technically, this is a series of successive merger
simulations. The first simulation includes G1 and G2 interacting at $z \sim
14.4$. It stops at $z \sim 13$ and a new galaxy G3 is added into the
system. During this process, all the dynamical properties of the pre-existing
system (e.g. G1 and G2 in this case) are preserved, while G3 is added based on
its properties and orbital parameters derived from cosmological
simulations. Then a second merger simulation with G1, G2 and G3 starts. Such a
procedure is repeated until all the progenitors enter the system. In the end,
the simulation includes all eight galaxies. Eventually all these galaxies
and black holes merge together. The duration of each merger simulation is
determined by the merger tree. The redshift at which each progenitor galaxy
enters the merger tree, the properties of each progenitor galaxy, and the
numerical parameters of the merger simulations are listed in Table~1. Below we 
describe the specification of these parameters.  

\subsubsection{Galaxy Models}
The structure of the galaxy models is motivated from leading theories of  
dissipational disk galaxy formation in CDM cosmologies that, as shown by
\cite{Mo1998}, are successful in reproducing the observed properties of both
present-day disk galaxies and damped Ly$\alpha$ absorbers in quasar spectra at
high redshift. The initial galaxy models are constructed in dynamical
equilibrium using a well-tested method (\citealt{Hernquist1993, Springel1999,
Springel2000, Springel2005B}). A halo is identified with a virial mass $\Mvir$
and a virial radius $\Rvir$ within which the overdensity $\Delta =
\rho_0/\rhoc = 200$, where $\rho_0$ and $\rhoc$ are the mean and critical
density, respectively. The density profile of the dark matter halo follows a
Hernquist profile \citep{Hernquist1990}, scaled to match that found in
cosmological simulations \citep{Navarro1997}, as described in
\cite{Springel2005B}:  

\begin{equation} 
\label{eq_hernquist}
\rho_{\rm H}(r) = \frac{\Mvir}{2\pi}\,\frac{a}{r(r+a)^3}\,, 
\end{equation}
where $a$ is a parameter that relates the \cite{Hernquist1990} profile parameters
to the appropriate NFW halo scale-length $R_{\rm s}$ and
concentration $\Cvir$ ($\Cvir = \Rvir /R_{\rm s}$),  
\begin{equation}
a = R_{\rm s}\sqrt{2 [ \ln(1+\Cvir) - \Cvir/(1+\Cvir)]}\,.
\end{equation}

The exponential disk of stars and gas are then constructed as in
\cite{Hernquist1993} and \cite{Springel2005B}. The
properties of the galaxy, including the virial mass $\Mvir$, virial radius
$\Rvir$ and halo concentration $\Cvir$ are scaled appropriately with redshift,
as described in \cite{Robertson2006A}. In particular, for a progenitor with
virial velocity $\Vvir$ at redshift $z$, $\Mvir$ and $\Rvir$ are calculated
following \cite{Mo1998}, while $\Cvir$ is adopted from \cite{Bullock2001}, as
briefly outlined below:    

\begin{eqnarray}
  \Mvir & = & \frac{\Vvir^{3}}{10GH(z)}\,, \\
  \Rvir & = & \frac{\Vvir}{10H(z)} \,, \\
  H(z)  & = & H_0\left[\Omega_{\Lambda} +(1-\Omega_{\Lambda}-\Omega_{\rm
    m})(1+z)^2+\Omega_{\rm m}(1+z)^3\right]^{1/2}\,, \\
  \Cvir & = & 9 \left[\frac{\Mvir}{M_0}\right]^{-0.13}
  \left(1+z\right)^{-1} \,.
\end{eqnarray}
\noindent
where $G$ is the gravitational constant, and $M_0 \sim 8 \times 10^{12}\,
h^{-1}\, \Msun$ is the linear collapse mass at the present epoch.  

 We assume a baryon fraction of
$f_{\rm{b}}=0.15$ for these high-redshift galaxies based on the WMAP1 result 
\citep{Spergel2003}. The gas fraction of each progenitor is
extrapolated from the results of semi-analytic models of galaxy formation 
\citep{Somerville2001}, with 100\% gas disks at $z \ge 10$ and 
90\%  at $10 > z \gtrsim 8$. The multiphase ISM is
envisioned to consist of cold clouds embedded in a hot, tenuous gas in
pressure equilibrium. Stars form out of the cold clouds by gravitational 
instability \citep{Li2005} with a rate that is proportional to the surface
density of the gas (\citealt{Schmidt1959, Kennicutt1998}; \citealt{Li2006B}).

In the adopted ISM model for the gas, the equation of state (EOS) is
controlled by a parameter $q_{\rm{EOS}}$ that linearly interpolates between
isothermal gas ($q_{\rm{EOS}} = 0$) and a strongly pressurized multiphase ISM
($q_{\rm{EOS}} = 1$). This EOS describes the dynamics of star-forming
gas and accounts for the consequences of stellar feedback on galactic
scales, and enables us to construct equilibrium disk models even with
large gas fractions \citep{Robertson2004, Springel2005E}.
Supernova feedback is modeled through thermal energy input
into surrounding gas and can help evaporate the cold clouds to
replenish the hot phase. For the simulation presented here a value of
$q_{\rm{EOS}} = 0.5$ is used, but test simulations using $q_{\rm{EOS}}
= 0.25-1.0$ produce average star formation and black hole accretion
rates that converge to within 15\%.

\subsubsection{Black Hole Accretion and Feedback}

The supermassive black holes are represented by collisionless ``sink''
particles. They interact with other particles gravitationally, and accrete
the gas. Accretion of gas onto the black holes is modeled using a
Bondi-Hoyle-Lyttleton parameterization (\citealt{Bondi1952, BondiHoyle1944,
  Hoyle1941}), in which the black holes accrete spherically from a stationary,
uniform distribution of gas, as described in \cite{DiMatteo2005} and
\cite{Springel2005B}: 

\begin{equation}
\label{eq_bondi}
\dot{M}_{\rm B} \, = \, {{4\pi \,
\alpha \, G^2 M_{\rm BH}^2 \, \rho} \over {(c_s^2 + v^2)^{3/2}}} \, ,
\end{equation}
\noindent where $M_{\rm {BH}}$ is the black hole mass, $\rho$ and $c_s$ are the 
density and sound speed of the gas, respectively, $\alpha$ is a dimensionless
parameter of order unity, and $v$ is the velocity of the black hole relative
to the gas.  

We assume the accretion has an upper limit by the Eddington rate,
\begin{equation}
\dot{M}_{\rm Edd} \, \equiv \, {{4\pi \, G \, M_{\rm BH} \, m_{\rm p}} \over
{\epsilon_{\rm r} \, \sigma_{\rm T} \, c}} \, . 
\end{equation}
\noindent where $m_{\rm p}$ is the proton mass, $\sigma_{\rm T}$ is the
Thomson cross-section, and $\epsilon_{\rm r}$ is the radiative
efficiency. The latter determines the conversion efficiency of mass accretion into
energy released as radiated luminosity. We adopt a fixed value of
$\epsilon_{\rm r} =0.1$, which is the mean value for radiatively
efficient \citet{Shakura1973} accretion onto a Schwarzschild
black hole.  In the simulations, the accretion rate is then the minimum of
these two rates, $\dot{M}_{\rm BH} \, = \, {\rm min} (\dot{M}_{\rm Edd}\, , \,
\dot{M}_{\rm B})$. 

The feedback from the black holes is associated with the mass accretion. We
assume that a small fraction ($\simeq 5\%$) of the radiated energy  
couples to the surrounding gas isotropically as feedback in form of thermal
energy. This fraction is a free parameter, determined by matching the observed
$\msigma$ relation \citep{DiMatteo2005}. For more discussions on this
prescription, see \cite{Hopkins2006A}. This feedback scheme self-regulates the 
growth of the black hole, and has been demonstrated to successfully reproduce
many observed properties of elliptical galaxies, as mentioned earlier.

\subsubsection{Black Hole Seeds}

To grow a black hole up to $10^9\, \Msun$ in less than 800 million years, a
wide range in seed masses, from $10 \, \Msun$ to $10^{6} \, \Msun$, have been
suggested (e.g., \citealt{Carr1984, Loeb1994, Bromm2003, Haiman2004, Yoo2004,
Volonteri2005, Begelman2006}). The formation of the black hole seeds
remains an open question, and several scenarios have been proposed. In
particular, \cite{Fryer2001} show that rapid collapse of massive PopIII stars
due to pair instability could produce black hole of $\sim 10^2\, \Msun$;
\cite{Bromm2003} suggest that hot and dense gas clump may collapse monolithicly to
form a massive black hole of $\sim 10^6\, \Msun$ in metal-free galaxies with a
virial temperature of $10^4$ K; while \cite{Begelman2006} propose that $\sim 20\,
\Msun $ black holes could form by direct collapse of self-gravitating gas due to
global instabilities in protogalactic halos, they then grow to $10^{4-6}\,
\Msun$ with super-Eddington accretion. We adopt the picture where
black hole seeds are the remnants of the first stars (\citealt{Abel2002,
Bromm2004, Tan2004, Yoshida2006, Gao2006}). The remnant black hole mass is
currently uncertain and widely debated.  Recent theory of PopIII star
formation predicts a mass range of $\sim 30 - 500\, \Msun$, but there are
two regimes where a SMBH could form, either $ \le 100 \, \Msun$ or $\ge 260\,
\Msun$ (\citealt{Heger2003}, see also \citealt{Yoshida2006} for a recent
discussion). We have tested the seed mass in the range of 100--300~$\Msun$
and find that the exponential growth of the black holes during the merger
makes our results insensitive to the choice in that range. We therefore assume
that the black hole seed starts with an initial mass of $200\, \Msun$ after
the collapse of the first star at $z=30$.    

These seed black holes then grow in the centers of $\sim 10^6\, \Msun$
halo which contains a large amount of high density  primordial gas, as current
theories predict that only one star forms per such mini-halo. The dense gas in
the central region provides abundant fuel for BH accretion. To account for
their evolution before the major mergers take place, the black holes
are assumed to grow at the Eddington rate until their host galaxies enter the
simulated merger tree. Such an approximation is supported by the
fact that the Eddington ratio in the simulations depends on the galaxy
interaction and strength of the feedback from the black holes. In our
simulations, most black holes grow at nearly the Eddington rate in the early
stages of a galaxy interaction when the feedback is weak. However, when the
interaction and the feedback become stronger, the Eddington ratios fluctuate
by orders of magnitude. So a constant accretion rate at the Eddington limit is
no longer appropriate, as we show below. Under this assumption, the first
progenitor galaxies (G1 and G2) of the quasar host have black hole seeds of
order $2\times 10^4\, h^{-1}\, \Msun$ by the time it enters the merger tree at
$z=14.4$. However, we should emphasize that this assumption serves only as an
upper limit of the early growth of the black holes. Our results in
the next sections imply that even if all the black hole seeds had a uniform
mass of $\sim 10^5\, \Msun$ when they enter the merger tree, it is still
possible to build a massive one to $10^9\, \Msun$ at $z \sim 6.5$ through
gas-rich mergers.

In our model, mergers are invoked in the formation of the most massive
black holes of $\gtrsim 10^7\, \Msun$ because that requires large supplies of
gas. Early on, however, this may not be necessary to grow the black hole seeds
from $\sim 100\, \Msun$ to the $\sim 10^5\, \Msun$ we start from, because the
accretion rate is small so other gas fueling could be sufficient. As
demonstrated in \cite{Hopkins2006E}, faint AGNs could be fueled by stochastic
accretion of cold gas that does not involve mergers. A similar process could
go on in the black hole seeds left by the PopIII stars at very high redshifts.
We should point out that in our simulations, it is necessary for galaxy 
progenitors in the merger tree to have reasonable massive black hole seeds
($\sim 10^5\, \Msun$) initially in order to build a $10^9\, \Msun$ black hole
at $z \sim 6.5$. However, our results are insensitive to specific
formation recipes of the seeds. The formation of seed black holes at high
redshifts is a challenging problem, and some of the proposed scenarios
mentioned above may indeed be necessary to make our seeds. However, currently
there is no observation available to test these models. 

In the picture we adopt in which the seeds come from the first stars, the
early accretion may be complicated by the feedback from the stars. We note that
recent studies by \cite{Johnson2006, Abel2006} and \cite{Yoshida2007} show
that HII regions form around the first stars, and that the halo gas would be
photo-ionized, photo-heated, and evacuated by the radiation feedback from the
stars. \cite{Johnson2006} suggest that such feedback would deplete the gas in
the central region, and would delay the black hole accretion by up to $10^8$
yrs. However, this destruction effect depends sensitively on the lifetime of
these massive stars, and more importantly on the environment which determines
both the gas density profile and gas replenish from inflow of the expelled gas
or neighboring halos. In the simulations presented in \cite{Johnson2006}, the
box size is only $100\, h^{-1}\, \kpc$, too small to contain the large scale 
gravitational potential and the large wavelength density modes that drive gas
infall, so the initial gas density is low and the destruction timescale is
long in this case. However, the quasar halo in our simulation resides in the
highest density peak in a volume of $1\, h^{-1}\, \Gpc^3$, where the halo
potential and gas density, as well we the accretion rate are much higher
\citep{Gao2006}. For a $200\, \Msun$ black hole, the accretion rate at
Eddington limit is only $10^{-6}\, \Msun\, \yr^{-1}$, which corresponds to the
Bondi accretion of molecular gas with a typical temperature of $\sim 100$ K at
density $\sim 10^2\, \rm{cm^{-3}}$, as implied from equation
(\ref{eq_bondi}). Such a density requirement is satisfied with the initial
conditions of our model. Therefore, the gas re-incorporation timescale in our
case may be substantially shorter than that estimated in \cite{Johnson2006}. 
We will investigate in a future project the growth and evolution of the early
black holes after the death of the first stars in such a cosmological
environment, using hydro-radiation simulations that include both radiative 
transfer and black hole accretion with ultra-high resolutions.

\subsubsection{Numerical Parameters of Merger Simulations}

The merger tree contains eight galaxies engaging in seven major
mergers at different times. For each merger event, the initial orbits
of the incoming progenitors are set to be parabolic, consistent with
the majority of the major mergers in our cosmological simulation and
with previous findings \citep{Khochfar2006}.  The orientation of each
merging galaxy is selected randomly. The initial separation between each
merging pair is set to $R_{0} = R_{\rm vir}$, where $R_{\rm vir}$ is the
virial radius of the incoming system, while the pericentric distance is chosen
as $R_{\rm p} = 0.5 R_{\rm d}$, where $R_{\rm d}$ is the radial disk scale
length of the incoming system.  We have tested different choices of $R_{\rm
p}$ and orientations, and found that the impact of these
parameters is minor because the orbital properties of the progenitors
change rapidly through interactions with the multiple galaxies in the
system.

Throughout the merger simulation, the mass and force resolutions are fixed for
each particle type, and the total initial particle number of $1.0\times 10^6$
results in particle masses of $m_{\rm{h}} = 1.1\times 10^7\, h^{-1}\,
\Msun$ for the halo and $m_{\rm{g, s}} = 2.2 \times 10^6\, h^{-1}\,
\Msun$ for both the gas and stars. The gravitational softening lengths are
$\epsilon_{\rm{h}} = 60\, h^{-1}\, \rm{pc}$ for halo particles and
$\epsilon_{\rm{g, s}} = 30\, h^{-1}\, \rm{pc}$ for both gas and stars. In the 
simulations, it is impossible to resolve individual stars, 
and the accretion radii of some small black holes are under-resolved.
However, with the sub-resolution implementation in our models, we can calculate
time-averaged rates of star formation and black hole accretion from the
large-scale properties of the gas, which are well resolved in our simulations.    
Resolution studies of a single merger \citep{Springel2005B} with particle
numbers from $1.6\times 10^5$ to $1.28 \times 10^7$ show that resolution affects
some fine structures of the gas and the instantaneous growth rates of star and
black holes, but the time-averaged properties of the system converge to within
20\%. 

\subsubsection{Halo Escape Velocity}
\label{subsec_Vesc}

\begin{figure}
\begin{center}
\includegraphics[width=3.5in]{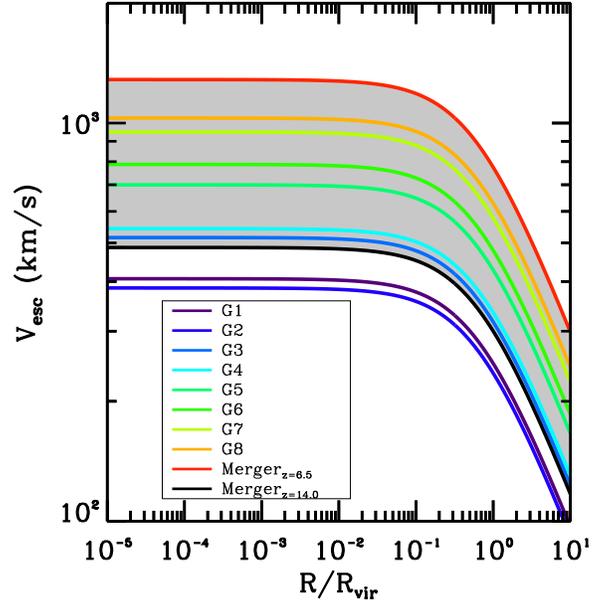}
\caption{
The halo escape velocity $V_{\rm esc}$ as a function of distance $R/\Rvir$
($\Rvir$ is the virial radius) to the halo center for various models in our
merger simulations. This plot includes the isolated halo 
progenitors G1 -- G8 in Table~1, as well as the first merger remnant at $z
\simeq 14$ and the last one at $z \simeq 6.5$, as labeled in the legend. The
shaded region indicates the range of the escape velocities of
the mergers in our simulations, with the values in the central regions being $486\,
\rm{km\, s}^{-1} \lesssim V_{\rm esc}  \lesssim 1284\, \rm{km\, s}^{-1}$. } 
\label{FigVesc}
\end{center}
\end{figure}

In a galaxy merger with black holes, the black holes may merge into one, or
may be ejected by gravitational recoil in the final stage. Their fate depends
on the halo escape velocity $V_{\rm esc}$. If the recoil velocity is larger
than $V_{\rm esc}$, then the black hole will be kicked out of the halo. We
follow \cite{Binney1987} to calculate this important parameter $V_{\rm
  esc}$. It is defined by
\begin{equation}
V_{\rm esc}(r) = \sqrt{2\left|\Phi(r)\right|}\,, 
\end{equation}
where $\Phi(r)$ is the gravitational potential at a given radius $r$. Because
the halo is spherical, the potentials of different spherical shells add
linearly, so $\Phi(r)$ is contributed by two parts, i.e., shells within $r$
($r' < r$) and outside ($r' > r$): 
\begin{equation}
\Phi(r) = -4\pi G\left[\frac{1}{r}\int^r_0 \rho_{\rm H}(r')\, r'^2\, \rm{d}r' + 
  \int^\infty_r \rho_{\rm H}(r')\, r'\, \rm{d}r'\right]\,.  
\end{equation}
where $\rho_{\rm H}(r)$ is again the \cite{Hernquist1990} density profile of
dark matter halo as in Equation~\ref{eq_hernquist}.  

Figure~\ref{FigVesc} shows the escape velocities of the halo
progenitors G1 -- G8 in Table~1, as well as two merger remnants at $z \simeq
14$ and $z \simeq 6.5$, respectively. The escape velocity depends on the halo
mass, redshift, and distance from the halo center. The $V_{\rm esc}$ remains
constant in the central region, begins to decline around $0.1\Rvir$. At the
center, $V_{\rm esc} \sim 2.5 \Vvir$, while at the virial radius
$\Rvir$, the escape velocity is comparable to the virial velocity (by a factor
of $\sim 1.5$). The isolated halo progenitors G1 -- G8 have a $V_{\rm esc}$
range of $\sim 385$ -- $1029\, \rm{km\, s}^{-1}$. The first merger halo at $z
\simeq 14$, which has a mass of $1.66 \times 10^{11}\, \Msun$ as the merger of
G1 and G2, has a central escape velocity of $V_{\rm esc} \sim 486\, \rm{km\,
s}^{-1}$, while the final merger halo at $z \simeq 6.5$, which has a mass of
$7.7 \times 10^{12}\, \Msun$, has $V_{\rm esc} \sim 1284\, \rm{km\,
  s}^{-1}$. The shaded region indicates the range of the halo escape
velocities of the mergers in our simulations. In particular, the escape speed
in the halo central region has a range of $486\, \rm{km\, s}^{-1} \lesssim
V_{\rm esc}  \lesssim 1284\, \rm{km\, s}^{-1}$. This range is important for
analysis of black hole ejection from gravitational recoil in the 
black hole binaries in \S~\ref{subsec_bhgrowth} and  
\S~\ref{subsec_bhbinary}.

\section{Formation of A Luminous $z \sim 6$ Quasar}

\subsection{Hierarchical Assembly of the Quasar Host}

\begin{figure*}
\begin{center}
\includegraphics[width=7in]{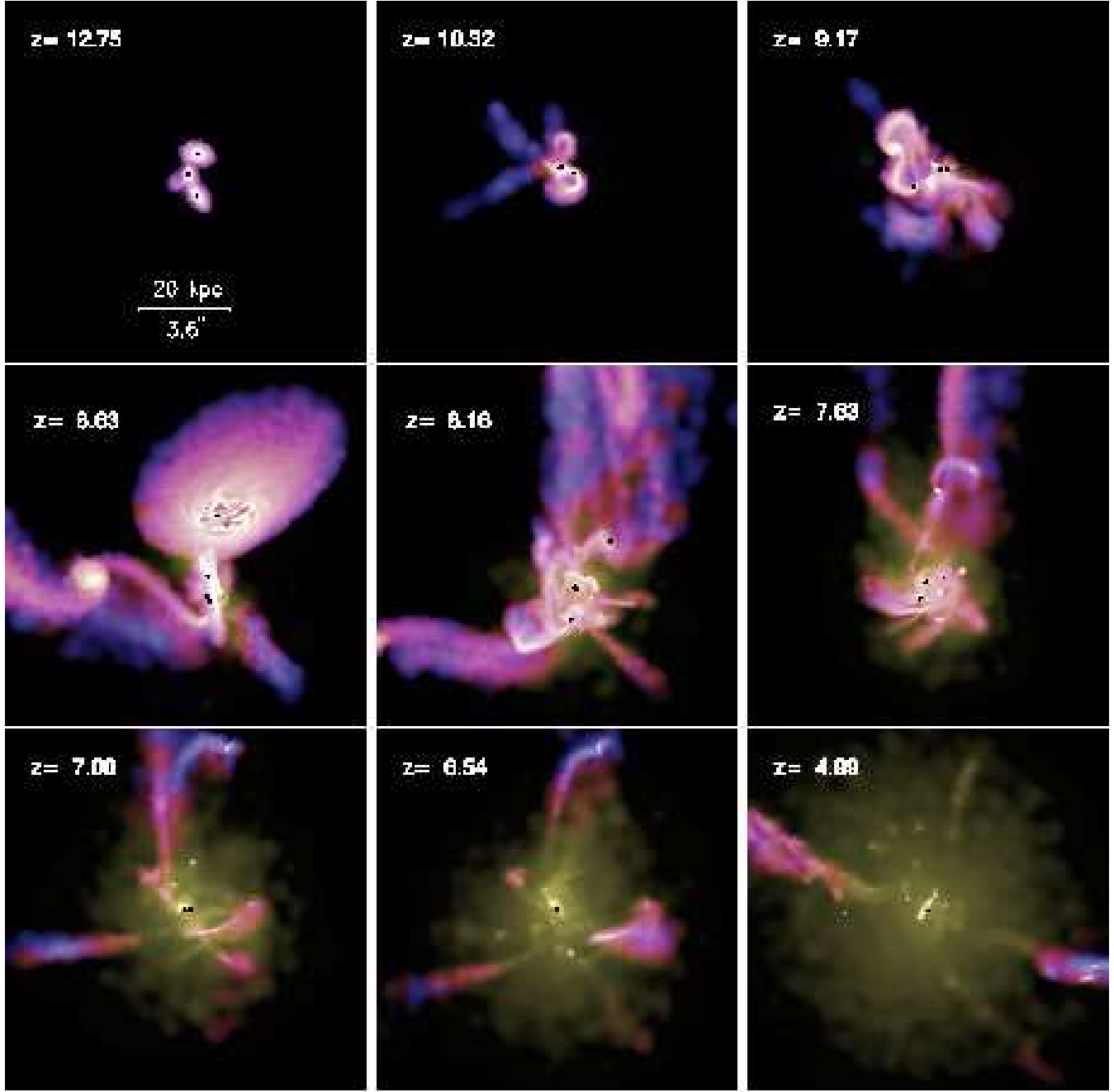} 
\vspace{0.5cm}
\caption{
  History of the quasar host shown in selected 
  snapshots. The images give the projected gas density, color-coded by
  temperature (blue indicates cold gas, yellow indicates hot, tenuous
  gas). The black dots represent black holes. There are eight galaxies in
  total, engaging in seven major mergers along the timeline of the 
  merger events as listed in Table~1. \textit{Top panels} show interactions in
  the early stage from $z \sim 13$ to 9. \textit{Middle panels} show the last
  major mergers between $z \sim\, $9--7, and \textit{bottom panels} show
  the final phase. All the galaxies coalesce at $z \approx 6.5$, creating an
  extremely luminous, optically visible quasar (see
  Figure~\ref{FigPic_Star}). At this time, there are three
  black holes, but the luminosity is dominated by the most massive one, which
  is more than two orders of magnitude larger than the others. These black
  holes merge into a single one at later time. The scale bar indicates a size
  of $20\, \kpc$ (comoving), corresponding to an angular size of $3.6''$ at
  redshift $z = 6.5$.}     
\label{FigPic_Gas}
\end{center}
\end{figure*}

\begin{figure*}
\begin{center}
\includegraphics[width=7in]{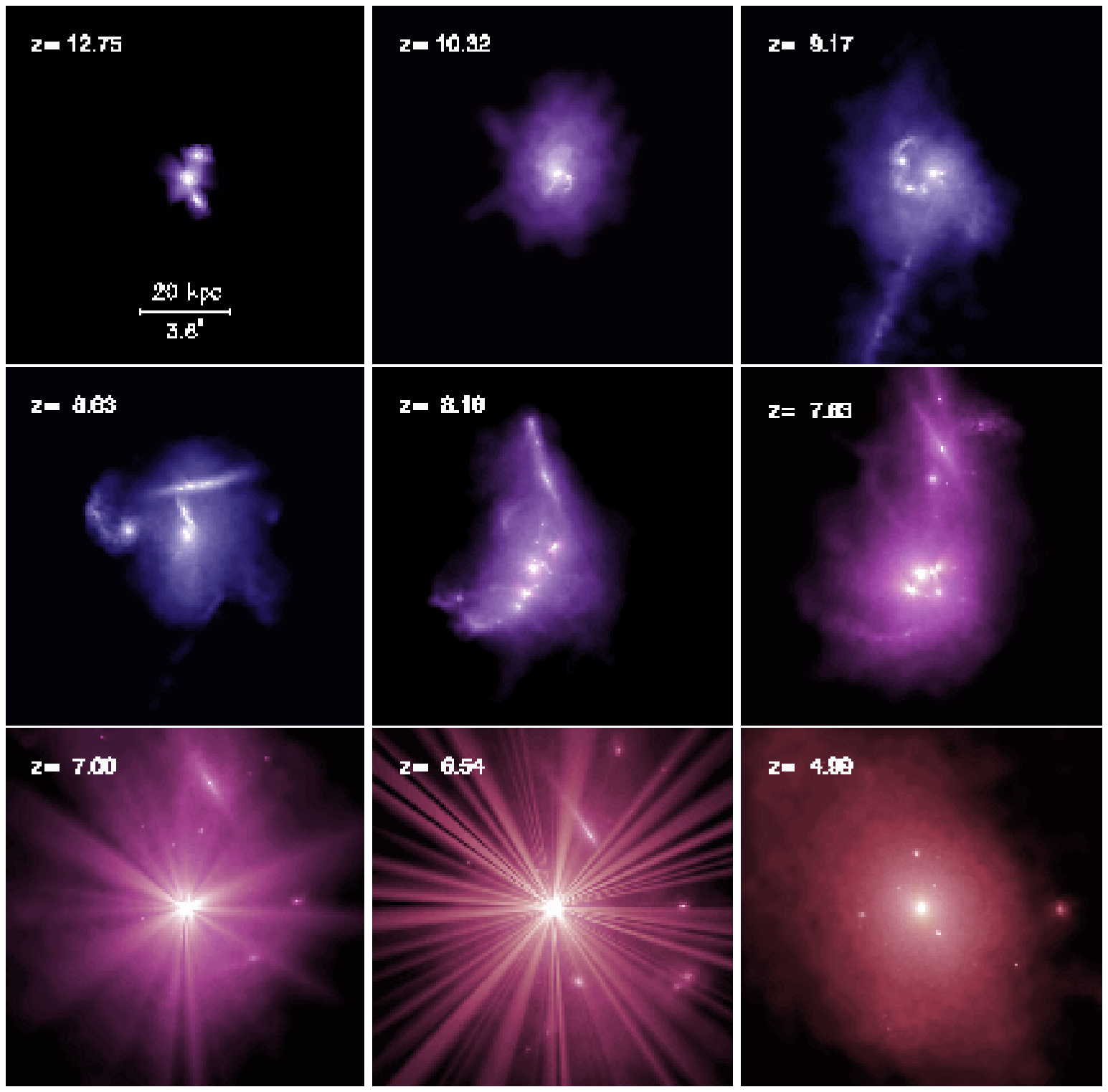} 
\vspace{0.5cm}
\caption{
  Same as Figure~\ref{FigPic_Gas}, but here the images
  show the projected stellar density, color-coded by the specific star
  formation rate (star formation rate per unit stellar mass). Blue indicates
  massive star formation in the galaxies, while red indicates little star
  formation. To illustrate the quasar activity, we have generated ``rays''
  around the quasar. The number and strength of the rays are proportional to the
  bolometric luminosity of the black holes. These rays are artificial and serve
  only as a visual guide. The systems in \textit{top panels} are blue, small
  and perturbed. The quasars appear very faint and buried. In the 
  \textit{middle panels}, strong interactions between galaxies boost star
  formation and black hole accretion, creating highly irregular
  morphologies and extremely blue galaxies. The quasars are heavily
  obscured by dense gas. At a later stage (\textit{bottom panels}), feedback
  from the black holes quenches star formation, allowing the galaxy color to
  redden. The quasar becomes optically visible as strong outflows blow out the
  gas. It has a maximum luminosity around $z \approx 6.5$ when all the galaxies
  coalesce. After that, both the quasar activity and star formation gradually
  die down, leaving behind an aging stellar spheroid.}  
\label{FigPic_Star}
\end{center}
\end{figure*}

The vigorous merging history of the quasar host is illustrated through selected
snapshots of the gas and stellar distributions in Figure~\ref{FigPic_Gas} and
Figure~\ref{FigPic_Star}, respectively. The progenitors at high redshifts are
very compact and gas rich. As the host galaxy of the quasar builds up
hierarchically, strong gravitational interactions between the merging galaxies
lead to tidal tails, strong shocks and efficient gas inflow that triggers
large-scale starbursts, a phenomenon that has been demonstrated by
many numerical simulations (e.g., \citealt{Hernquist1989B, Hernquist1989A,
  Barnes1991, Barnes1996, Mihos1994, Mihos1996, Springel2000, Barnes2002,
  Naab2003, Li2004}), as reviewed by \cite{Barnes1992}. The highly
concentrated gas fuels rapid accretion onto the SMBHs (\citealt{DiMatteo2005,
  Springel2005B}). Between $z \sim$~14--9, the merging systems are
physically small and the interactions occur on the scale of tens of
kiloparsecs. By $z \sim$~9--7, when the last major mergers take place, the
scale and strength of interactions have increased dramatically. Galaxies are
largely disrupted in close encounters, tidal tails of gas and stars extend
over hundreds of kiloparsecs, and intense bursts of star formation are
triggered.

The black holes continue to grow rapidly during this period but are heavily
obscured by a significant amount of circumnuclear gas. During galaxy mergers, the
black holes follow their hosts to the center of the system and can interact
closely with each other. It has been shown that black hole binaries decay rapidly
in a gaseous environment and can merge within $\sim 10^7$ yrs
(\citealt{Escala2004, Li2007C}). Because the galaxies in our simulations are
very gas rich and the gas is highly concentrated during the mergers,
we therefore assume that the black holes merge efficiently owing to strong
dynamical friction with the gas \citep{Springel2005B}. We will return to
more discussions of this process in \S~\ref{subsec_bhgrowth} and
\S~\ref{subsec_bhbinary}.     

At redshift $z \approx 6.5$ the progenitor galaxies coalesce, inducing high
central gas densities that bring the SMBH accretion and feedback to a
climax. The SMBH feedback then drives a powerful galactic wind that clears the 
obscuring material from the center of the system.  The largest SMBH becomes
visible as an optically-bright quasar \citep{Hopkins2006A} during this phase,
after which quasar feedback quenches star formation and self-regulates SMBH
accretion. Consequently, both star formation and quasar activity die down,
leaving a remnant which reddens rapidly, as illustrated schematically in
Figure~\ref{FigPic_Star}.  

\subsection{Star Formation History}

\begin{figure}
\begin{center}
\includegraphics[width=3.5in]{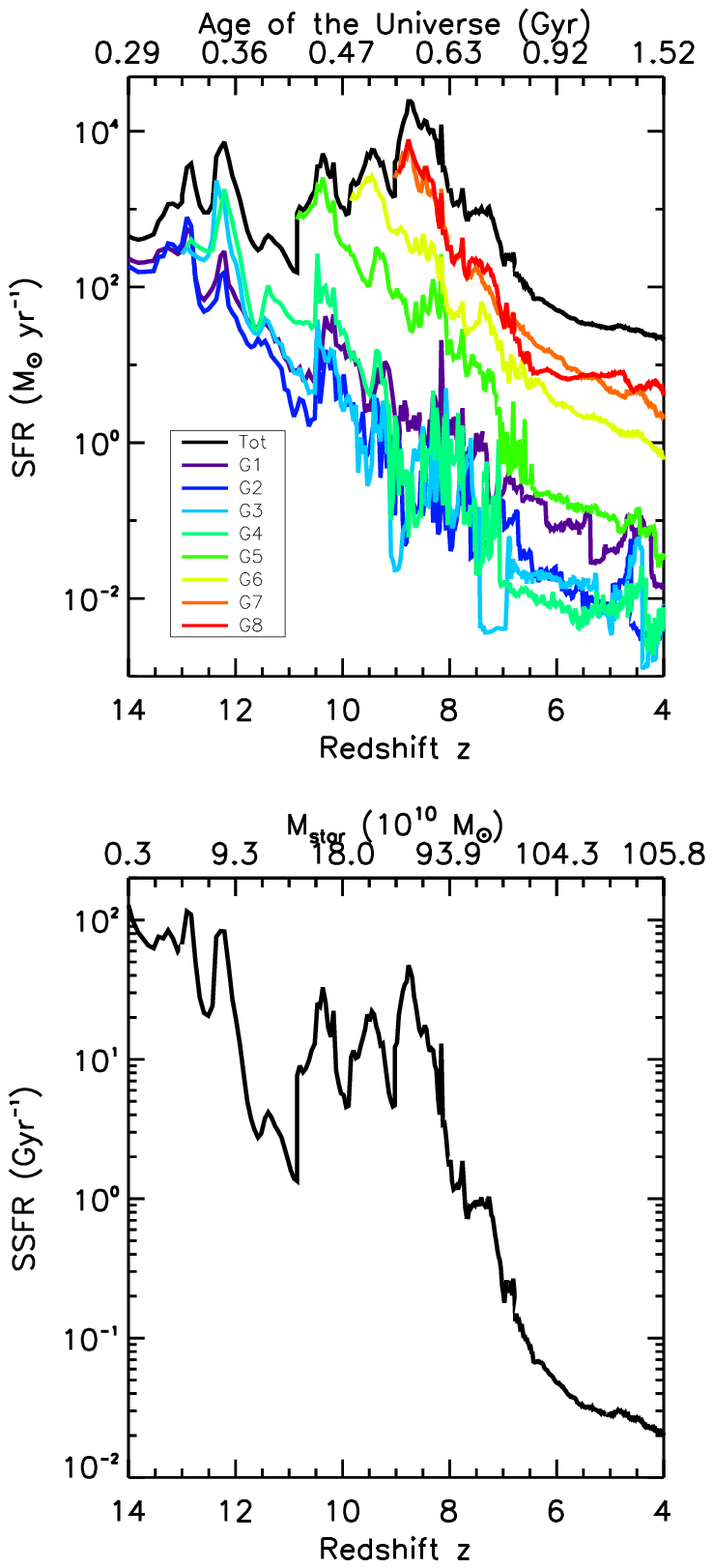}
\caption{Time evolution of star formation rate (SFR, {\em top panel}) and 
  specific star formation rate (SSFR, {\em bottom panel}), respectively. The
  colored lines indicate individual galaxies, while the black lines give
  summed quantities for the entire system.}   
\label{FigSFR}
\end{center}
\end{figure}

The evolution of the star formation rates (SFRs) of each individual galaxy, and 
total SFR of the whole system are shown in Figure~\ref{FigSFR} ({\em top
  panel}). The system forms stars rapidly as these compact and gas rich
progenitors undergo strong interactions. The total SFR ranges from $\sim 100
\, \Msun\, \yr^{-1}$ to $> 10^{4}\, \Msun\, \yr^{-1}$ between redshifts $z
\sim$~9--8 when the galaxies begin their final major mergers, while the SFRs of 
individual galaxies fall below a few$\times 10^{3}\, \Msun\, \yr^{-1}$, within
the starburst intensity limit of $10^3 \, \Msun\, \yr^{-1}\, \kpc^{-2}$
proposed by \cite{Meurer1997} and \cite{Thompson2005}. At $z<7$ the star
formation rate decreases gradually owing to a depletion of the gas supply and
progressively stronger feedback from the SMBHs. At the time of final
coalescence ($z \approx 6.5$) the star formation rate is $\sim100 \, \Msun\,
\yr^{-1}$, an order of magnitude lower than for estimates of \zquasar\
(\citealt{Bertoldi2003A, Carilli2004}). We note, however, that the estimates by
these authors are based on the assumption that the FIR luminosity is dominated
by young stars, and they cannot rule out the possibility that AGN may
contribute significantly to the luminosity.

In a forthcoming paper \citep{Li2007A}, we have calculated the infrared
properties of the quasar system using a 3-D Monte Carlo radiative transfer
code that incorporates adaptive grids and treats dust emission
self-consistently. We find that the far-infrared luminosity of our quasar is
not dominated by young stars but instead has a substantial quasar contribution
of over $80\%$. This finding is supported by observations of \zquasar\ in
near-IR (e.g., \citealt{Charmandaris2004, Hines2006}, and more recently
\citealt{Dwek2006}), which show a remarkably flat spectral energy distribution
and suggest an AGN origin for the flux excess. Furthermore, adopting a total
gas mass of $\sim 10^{10}\, \Msun$ (\citealt{Walter2004, Narayanan2006a}) in
\zquasar, a simple application of the Schmidt-Kennicutt star formation law
(\citealt{Schmidt1959, Kennicutt1998}) gives a star formation rate of $\sim
200\, \Msun\, \yr^{-1}$, close to what we find here.

Within only about 600 Myrs from $z=14.4$ to $z=6.5$, the system accumulates a
stellar mass of $\sim 10^{12}\, \Msun$ as shown in Figure~\ref{FigSFR} ({\em
  bottom panel}). The specific star formation rate (SSFR), or ``b-parameter''
(e.g., \citealt{Brinchmann2004}), is defined as $\rm {SSFR=SFR/M_{
  star}}$. It is a measure of the fraction of the total stellar mass currently
forming at a specific time. The SSFR is typically larger in high-redshift
galaxies than in ones at low redshifts owing to vigorous star
formation. During the past several years, there has been rapid progress in
observing galaxies at $z \gtrsim 6$ using the \textit{Hubble Space Telescope}
(HST), and the \textit{Spitzer Space Telescope} (Spitzer) coupled with
ground-based observatories (e.g.,
\citealt{Dickinson2004, Bunker2004, Bouwens2004, Giavalisco2004, 
  Egami2005, Eyles2005, Mobasher2005, Yan2005, Yan2006, Eyles2006}), and
hundreds of these distant objects have been detected. These frontier
observations suggest that the Universe experienced
rapid star formation during the redshift
interval $14 \gtrsim z \gtrsim 6$, and the development of 
large stellar systems in the mass range of $\sim$ $10^{10 - 11}\, \Msun$. In
particular, several groups (\citealt{Egami2005, Yan2006, Eyles2006}) find
SSFRs in the range of $10^{-1}$--$10^{2} \rm{Gyr}^{-1}$ in their
observations, consistent with our simulations.

\subsection{Metal Enrichment}

\begin{figure}
\begin{center}
\includegraphics[width=3.5in]{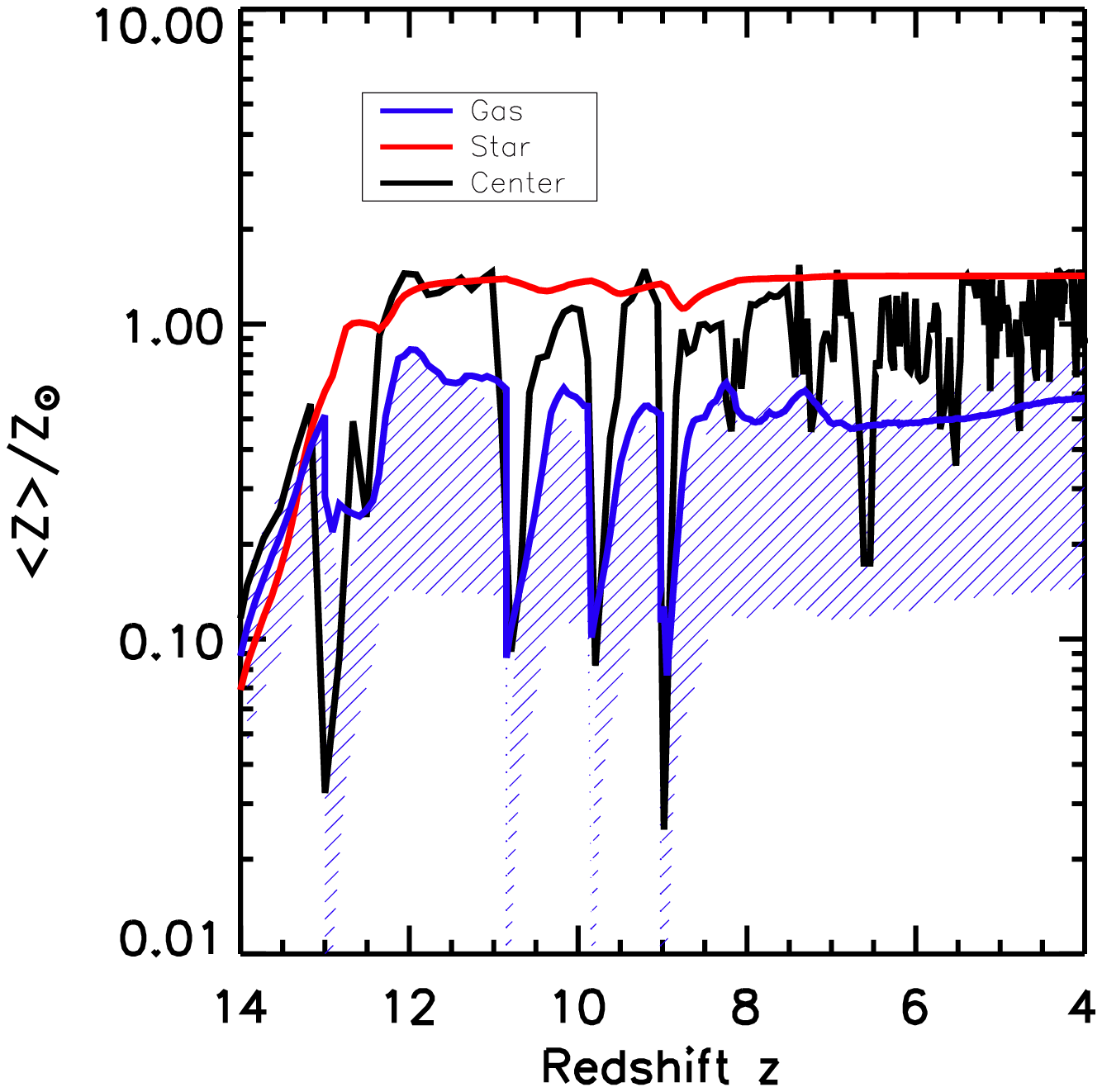}
\caption{Time evolution of mass-weighted metallicity in the quasar host, from
  gas (blue curve), stars (red curve), and the mean value in the central
  region ($R < 1\, \kpc$) of each galaxy (black curve). The blue hatched region
  indicates the range of 25\%--75\% of the gas metallicity. }  
\label{FigMetT}
\end{center}
\end{figure}

\begin{figure}
\begin{center}
\includegraphics[width=3.5in]{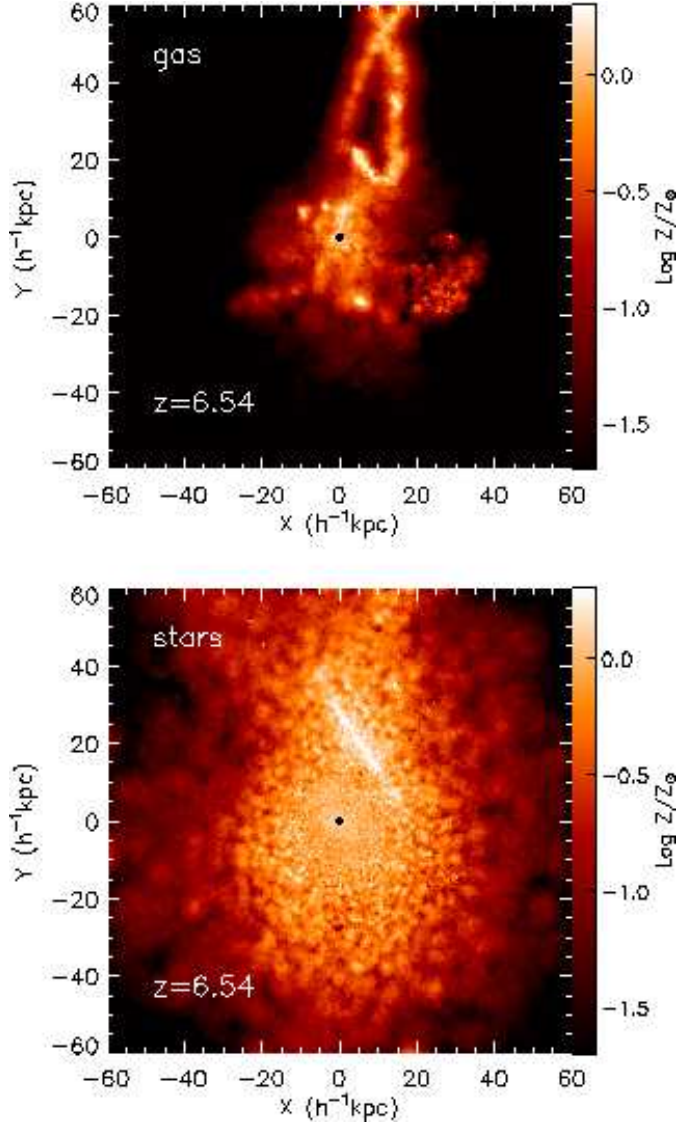}
\caption{Spatial distribution of mass-weighted metallicity of the quasar host
  at $z\approx 6.5$, from both gas ({\em top panel}) and stars ({\em bottom
  panel}), respectively. The images are projected metallicity adaptively
  smoothed over 32  particles (analog to the SPH kernel in a 2-dimensional
  plane). The black dot indicates the center of the quasar.}    
\label{FigMetR}
\end{center}
\end{figure}

Rapid star formation in the quasar progenitors produces an abundant 
mass of heavy elements to enrich the ISM. Observations of \zquasar\ show solar
metallicity in the system (\citealt{Barth2003, Walter2003, Maiolino2005,
Becker2006}). Figure~\ref{FigMetT} shows the metallicity in our simulated
quasar system at different times. Note that the dips and jumps in the curves
owe to incoming new galaxies which bring in metal-poor pristine gas and
newly-formed stars. The quasar host reaches solar metallicity as early as $z\sim
12$, and maintains similar levels to later times. The spatial distribution of
metallicity from both gas and stars at the peak quasar phase at $z \approx
6.5$ is shown in Figure~\ref{FigMetR}. The metals are widely spread owing to
outflow from the quasar feedback and gas infall toward the merger center. 
The metallicity in the central region of the merger remnant is slightly above
the solar value. In some outer regions, because the gas and stars are still
falling back to the system center, the infalling material triggers small-scale 
bursts of star formation. So the metallicity in these blobs appears to be 
super-solar, as shown in Figure~\ref{FigMetR}.

Calculations of carbon monoxide emission using non-local thermodynamic
equilibrium radiative transfer codes \citep{Narayanan2006b, Narayanan2006c}
by \cite{Narayanan2006a} show CO luminosities, excitation patterns, and
morphologies within the central $\sim 2\, \kpc$ of the quasar host center that
are consistent with observations of \zquasar\ (\citealt{Walter2003,
Bertoldi2003B, Walter2004}). These results were derived using Galactic CO
abundances, and thus support our conclusions that significant metal enrichment
takes place early in the quasar host, as a result of strong star formation in
the progenitors. 

\subsection{Growth of Supermassive Black Holes}
\label{subsec_bhgrowth}

\begin{figure}
\begin{center}
\includegraphics[width=3.5in]{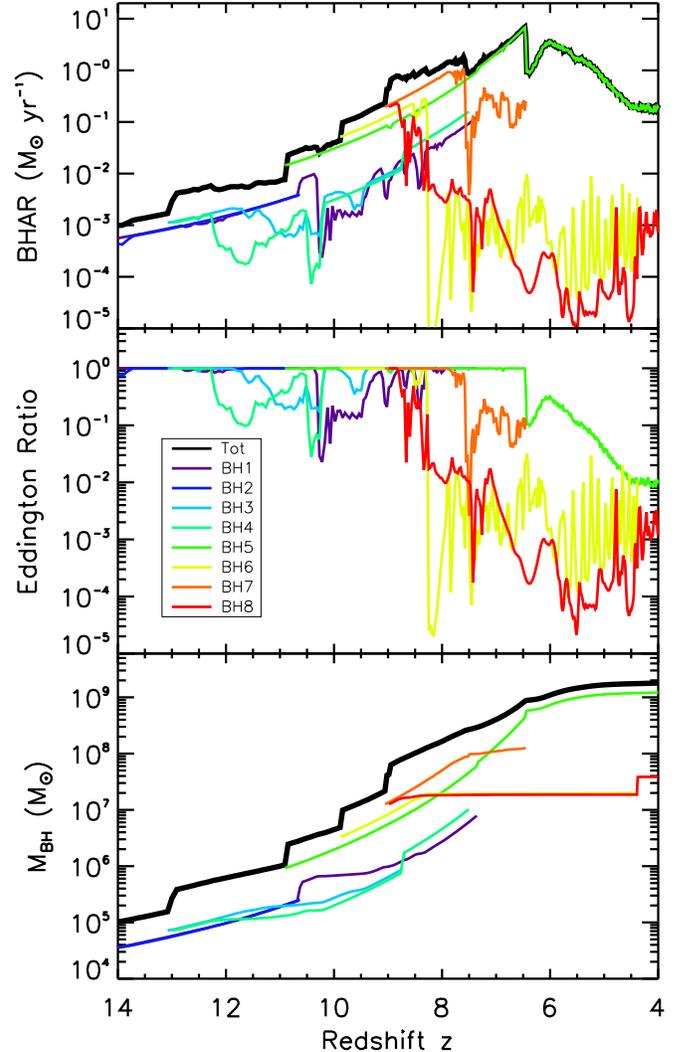}
\caption{The growth history of the quasar system, including the black hole
  accretion rate {BHAR, \em top panel}, the Eddington ratio $\Lbol/\Ledd$ 
  ({\em middle Panel}), and black hole mass ({\em bottom panel}). Note 
  that the
  black curves represent totals, while colored curves show
  individual black holes, as indicated in the legend.}  
\label{FigAR}
\end{center}
\end{figure}

In the simulations, the quasar host at $z \sim 6$ is built up by eight
progenitors, each containing a black hole in the center.
Figure~\ref{FigAR} shows the evolution of the black hole accretion rate, the
Eddington ratio, and the integrated masses of the whole system and individual
black holes. The total black hole accretion rate grows steadily during the
hierarchical assembly of the host galaxy and peaks at $\sim 10\, \Msun\,
\yr^{-1}$ around $z \approx 6.5$ during the final coalescence.

The Eddington ratio, $\Lbol/\Ledd$, of each individual black hole
varies with time, depending on the galaxy interaction and feedback from the
black holes. The black holes maintain accretion at the Eddington limit for
only a fraction ($<50\%$) of the time. At the peak of quasar activity,
the Eddington ratio of the most massive black hole is near unity, while that
of the other black hole is only 0.1. However, collectively, the whole system
appears to accrete at $\Lbol/\Ledd \sim 1$ at $z \gtrsim 6.5$, as implied in
Figure~\ref{FigAR}. Studies of black hole accretion (e.g.,
\citealt{Vestergaard2004, Kollmeier2006, Vestergaard2006A}) show that the
Eddington ratio has a wide range of 0.01--1.0, and it varies with both
luminosity and redshift. Luminous systems tend to have higher $\Lbol/\Ledd$
than less-luminous counterparts, and at $z \gtrsim 4$, most quasars shine at
nearly Eddington luminosity. Our results suggest that {\em individual} black
holes do not always necessarily accrete at the Eddington rate. 
However, since high-redshift, luminous quasars may form through mergers of
several galaxy progenitors containing black holes as in our case, therefore
the growth of the quasar represents a {\em collective} contribution from each
individual black hole. The total black hole mass increases from $\sim 6\times
10^4\, \Msun$ at $z\approx 14$ to about $2\times 10^9\, \Msun$ at $z \approx
6.5$, close to that estimated for \zquasar\ by \cite{Willott2003} and
\cite{Barth2003}. 

In the simulations, we do not have sufficient resolution nor the relativistic
physics to consider the ejection of black holes by gravitational recoil. The
black holes are assumed to merge efficiently once their separation is below
the spatial resolution. In the final stage of black hole mergers, the emission
of gravitational wave carries linear momentum, which could cause the black
holes to recoil (e.g., \citealt{Bonnor1961, Peres1962}). If the recoil
velocity is larger than the halo escape velocity, then the black holes will be 
kicked out from their halo (e.g., \citealt{Fitchett1983, Favata2004, 
  Merritt2004, Madau2004, Haiman2004, Yoo2004, Volonteri2005}). Previous
studies by \citep{Haiman2004, Yoo2004, Volonteri2005, Haiman2006} suggest that
constant or super-Eddington accretion is required to produce $10^9\, \Msun$
black holes at $z\sim 6$ if ejection of black holes is included. In
particular, \cite{Haiman2004} suggests a black hole will be ejected if 
the kick velocity $V_{\rm {kick}}$ for the coalescing SMBH binary is larger
than twice the halo velocity dispersion $\sigma_{\rm {halo}}$, $V_{\rm {kick}}
\gtrsim 2 \sigma_{\rm {halo}}$, as the dynamical friction timescale
for the kicked black hole to return to the halo center is longer than the
Hubble time \citep{Madau2004}. By applying this ejection criterion 
to a Press-Schechter merger tree of a  $8.5 \times 10^{12}\, \Msun$ halo
within which \zquasar\, is assumed to reside, \cite{Haiman2004} finds that the
SMBH of the quasar gains most of its mass rapidly from seed holes during $17
\lesssim z \lesssim 18$ due to black hole ejection, and the SMBH likely
accretes with super-Eddington rate in order to build a mass as that of
\zquasar.    

However, the halo escape velocities or velocity dispersions in
our model are much larger than the currently best estimates of the kick
velocity. The quasar halo in our simulations has active merging history from
redshifts $z \simeq 14.4$ to $z \simeq 6.5$, and the halo progenitors have
masses much higher than those considered in the previous studies
\citep{Haiman2004, Yoo2004, Volonteri2005}. The quasar halo builds its mass
from $\sim 1.16 \times 10^{11}\, \Msun$ at $z \simeq 14$ (the sum of
progenitors G1 and G2 in Table~1) to $7.7 \times 10^{12}\, \Msun$ at
$z=6.5$. As shown in Figure~\ref{FigVesc} (the shaded region), the central
escape velocity of the mergers in our simulations is in the range $486\, -
1284\, \rm{km\, s}^{-1}$. Currently, the maximum kick velocity for 
unequal-mass, non-rotating BH binary is in the range of $\sim 74$ -- $250\, 
\rm{km\, s}^{-1}$ from both analytic post-Newtonian approximation (e.g.,
\citealt{Blanchet2005, Damour2006}) and the ground-breaking full relativistic
numerical simulations (e.g., \citealt{Herrmann2006, Baker2006,
  Gonzalez2006}). For equal-mass, spinning BH binary, \cite{Favata2004} 
estimate a range of $\sim 100$ -- $200\, \rm{km\, s}^{-1}$ using 
BH perturbation theory, and \cite{Herrmann2007} derive
a formula from relativistic simulations, $V_{\rm {kick}} = 475S\,
\rm{km\, s}^{-1}$, where $S \le 1$ is the BH spin. This gives a maximum kick
of $475\, \rm{km\, s}^{-1}$ for maximal spin. Although it is also reported
that the recoil velocity can be as large as thousands $\rm{km\, s}^{-1}$
\citep{Gonzalez2007, Campanelli2007} for BH binary in the orbital plane with
opposite-directed spin. However, as pointed out by \cite{Bogdanovic2007}, such
a configuration is rather uncommon, especially in gas-rich galaxy mergers,
because torques from accreting gas suffice to align the orbit and spins of
both black holes with the large-scale gas flow. The resulting maximum kick
velocity from such a configuration is $< 200\, \rm{km\, s}^{-1}$. Overall,
the kick velocity from the latest calculations of black hole binary is in
the range of $\sim 100$ -- $475 \rm{km\, s}^{-1}$, falling safely below the
escape velocities of the quasar halos in our simulations, so black hole
ejection may be insignificant in our case. 

Moreover, we find that our model can produce a $10^9\, \Msun$ SMBH
even if ejection is allowed. From Figure~\ref{FigAR} (bottom panel), the
$10^9\, \Msun$ SMBH is dominated by BH5, most of its mass comes from gas
accretion. Even if the less massive black holes, for example BH7 or BH8 was
ejected, the most massive one BH5 is still able to reach $\sim 10^9\, \Msun$
in the end. Furthermore, even if all the seeds started with $\sim 10^5\,
\Msun$ in the merger tree, the result would be about the same. We therefore
conclude that the results from our modeling are robust. Suppermassive black
holes of $\sim 10^9\, \Msun$ can grow rapidly through gas accretion and
mergers hierarchically in the early Universe, constant or super-Eddington
accretion is not necessary, unless the recoil velocity of the coalescing black
hole binary is extremely high such that most of the black hole seeds in our
simulations are ejected  (e.g., $V_{\rm {kick}} > 1000\, \rm{km\, s}^{-1}$).

\subsection{Correlations between Supermassive Black Hole and Host Galaxy}

\begin{figure}
\begin{center}
\includegraphics[width=3.5in]{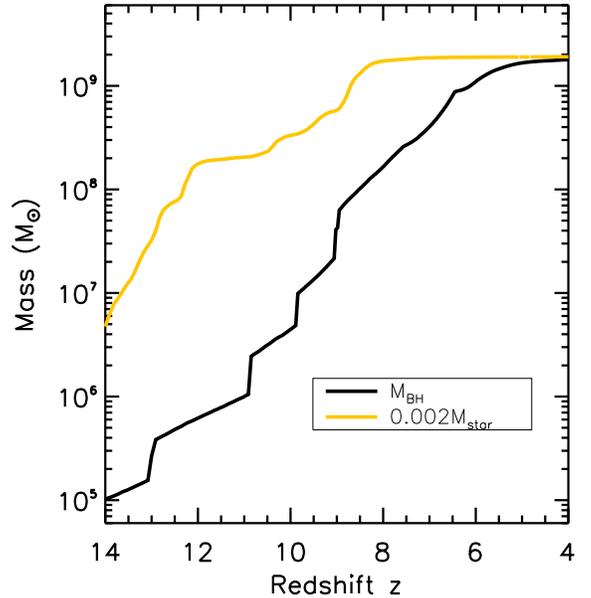}
\caption{Evolution of total BH (black curve) and stellar (yellow
  curve) mass, respectively. The stellar mass is multiplied by a factor of
  0.002, to reflect the observed correlation of $\MBH \approx 0.002\, \Mstar$
  at the present day, as parameterized by \cite{Marconi2003}.}  
\label{FigMas}
\end{center}
\end{figure}

Tight correlations between supermassive black holes and hosts have
been observed in local galaxies (e.g., \citealt{Magorrian1998, Ferrarese2000,
  Gebhardt2000}), but the inference of these relationships at higher redshifts
remains an open question.  Because the eight galaxies in our system  
interact vigorously with each other, the stellar components are widely spread
and mixed, it is impossible to separate individual galaxy-SMBH pairs, so
we only consider the correlations in {\em total} quantity of the whole system.
A comparison of the total stellar mass and total black hole mass is
shown in Figure~\ref{FigMas}. At early time, both the stars and black holes
grow rapidly through galaxy mergers. Shortly after the peak quasar phase, 
strong feedback suppresses both the accretion and star formation, the masses
of the black holes and stars become saturated gradually, and in the end  
satisfy $\MBH \approx 0.002\, \Mstar$, similar to the correlation measured in
nearby galaxies (\citealt{Magorrian1998, Marconi2003}). Our results
are consistent with findings by \cite{Robertson2006A} and \cite{Hopkins2007},
and demonstrate that the observed $\mbulge$ scaling relation is a result of
the coeval growth of the SMBH and its host galaxy, and that it holds across
different cosmic times.  

We note, however, that the velocity dispersion of the stars in the remnant
center is about $\sim 500\, \rm{km\, s}^{-1}$ (after the system relaxes) owing
to the deep potential well of the merger system, so the $\msigma$ relation
falls below the correlation observed locally (\citealt{Ferrarese2000,
  Gebhardt2000, Tremaine2002}). Single mergers of progenitor galaxies
constructed in a redshift range of z=0--6 by \cite{Robertson2006A} appear to
follow the observed $\msigma$ correlation with a weak redshift dependence of the
normalization, which results from an increasing velocity dispersion of the
progenitors at higher redshift. The multiple mergers we derive from
cosmological simulations take place at much higher redshifts and hence the
progenitors have larger velocity dispersions, implying a larger deviation from  
the local $\msigma$ relation than in the work of \cite{Robertson2006A}.
However, because we do not follow subsequent mergers and accretion into the
host halo below $z\sim 6$, the implications of this result for the evolution
of the $\msigma$ relation are unclear. 

Observations of active galaxies have yielded ambiguous results about the
SMBH--spheroid relationship. For example, \cite{Greene2006} report a lower
zero-point of the $\msigma$ of local active galaxies than that of the inactive 
sample \citep{Tremaine2002}; at $z>0$, \cite{Shields2003} found the 
same $\msigma$ relation in the redshift range $z=$1--3, while others (e.g.,
\citealt{Treu2004, Walter2004, Borys2005, Peng2006, Shields2006}) show
correlations with various offsets. In particular, \cite{Walter2004} estimate a
dynamical mass of $\sim 5 \times 10^{10}\, \Msun$ using the CO linewidth
measured in \zquasar, and suggest that the bulge is under-massive by at least
one order of magnitude compared to the local $\mbulge$ relation. However, the
CO calculation by \cite{Narayanan2006a} finds that the CO linewidth of the
quasar in our simulation is larger than the mean 280 km s$^{-1}$ measured by
\cite{Bertoldi2003B} and \cite{Walter2004} by almost an order of magnitude, 
and that the derived dynamical mass is $\sim 10^{12} \, \Msun$, putting the
simulated quasar on the $\mbulge$ correlation. \cite{Narayanan2006a} further
suggest that the observed emission line may be sitting on top of a much
broader line, which may be tested by future observations with large
bandwidths.  

The different relations reported from the observations may reflect a divergence
of the methods used to estimate the black hole mass and stellar properties, or
may represent different evolutionary stages of the systems \citep{Wu2006,
Hopkins2007}. More observations and better measurements of black hole mass and
properties of host bulges will be crucial to study the SMBH--host relations in
high-redshift quasar systems \citep{Vestergaard2006B} and to test our hypothesis. 

\subsection{Quasar Luminosities}

\begin{figure*}
\begin{center}
\includegraphics[width=6.0in]{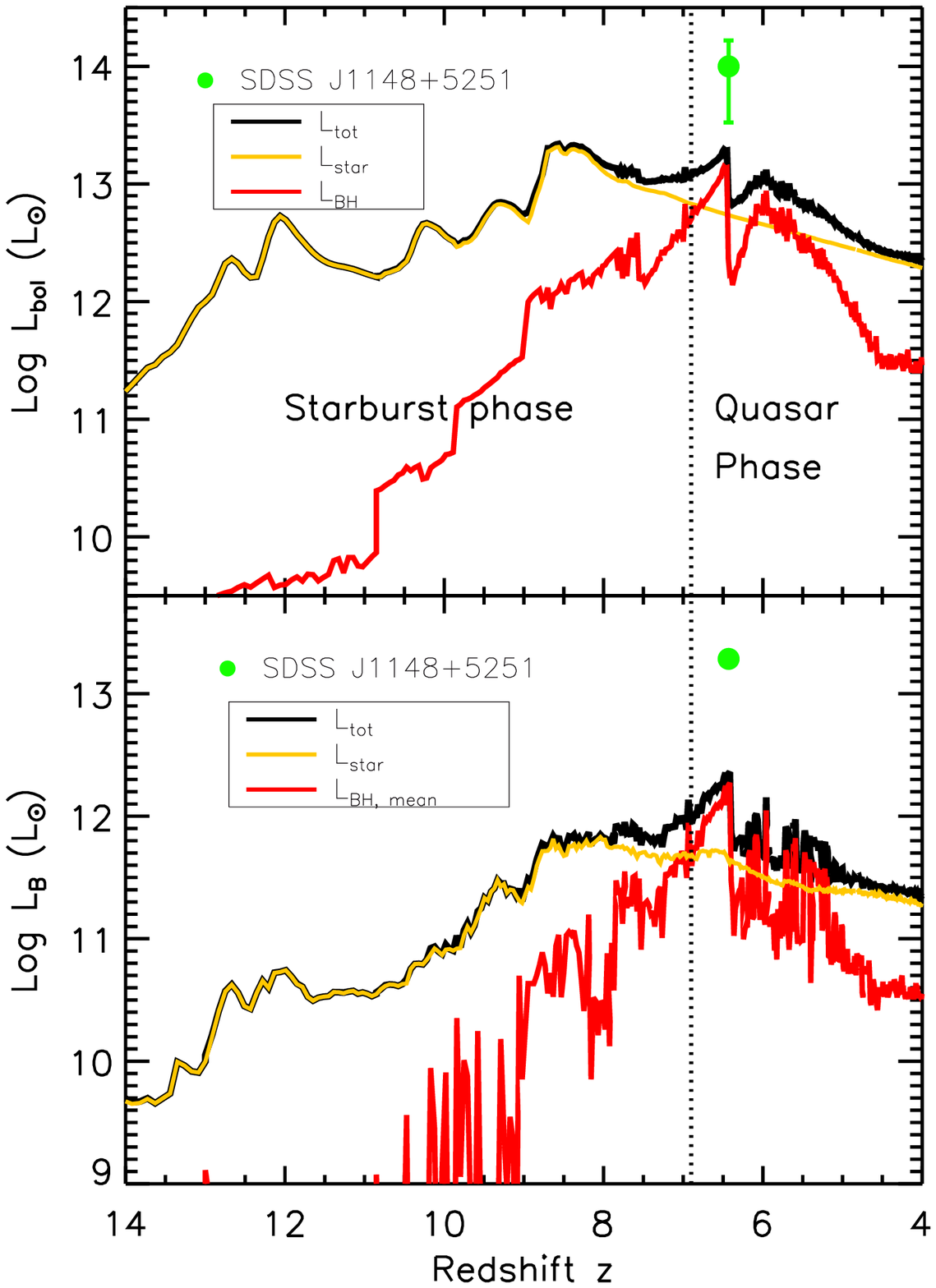}
\vspace{0.5cm}
\caption{Comparison of luminosities from simulations and observations. Shown
  are bolometric (\textit{top panel}) and attenuated luminosities in the
  rest-frame B-band (\textit{bottom panel}). Note $\Lbol$ of
  SDSS J1148+5251 (green dot) is an estimate for a SMBH of $3\times 10^9\, \Msun$
  accreting at the Eddington rate, with the error bar indicating the mass range
  of 1--5$\times10^9 \, \Msun$ \citep{Willott2003, Barth2003}, while the $\LB$
  is converted from observations at wavelength 1450$\AA$\ \citep{Fan2003}. The
  yellow, red, and black curves represent luminosities of stars, black
  holes, and total (sum of the above two), respectively. For the black holes, 
  $L_{\rm {BH, mean}}$ is the average luminosity over 1000 sight lines. 
  Note in the luminosity calculation, the black holes are assumed to be
  non-rotating. If the black holes are rotating, their radiative luminosities
  could be higher by up to a factor of a few, see text for more discussions.}       
\label{FigLum}
\end{center}
\end{figure*}

\begin{figure}
\begin{center}
\includegraphics[width=3.5in]{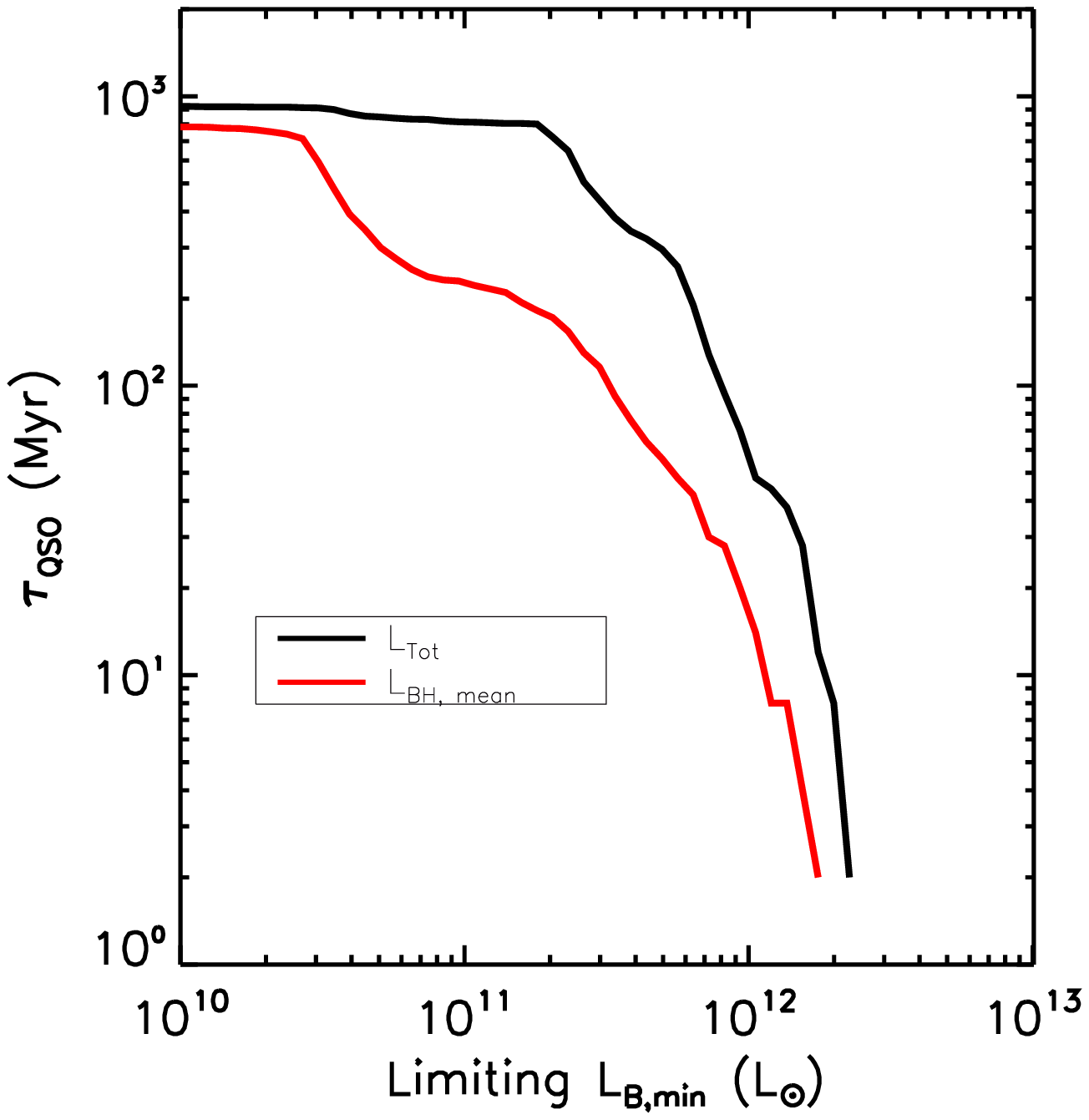}
\caption{Quasar lifetimes as functions of different B-band limiting
  luminosities. The black and red curves represent the total luminosity of
  the system, and mean luminosity over 1000 sight lines of the black
  hole, respectively, corresponding to the curves of the same color in the
  bottom panel of Figure~\ref{FigLum}.}  
\label{FigTq}
\end{center}
\end{figure}

Both the bolometric and attenuated luminosities of the quasar and the host
galaxy in the simulations can be readily calculated following the methodology
of \cite{Hopkins2005A}. The bolometric luminosity $\Lbol$ of stars is
calculated using the stellar population synthesis model of
\cite{Bruzual2003}, while that of a black hole is calculated as
$\Lbol=\epsilon_{\rm r}{\dot {M} \, c^2}$, where $\epsilon_{\rm r}=0.1$ is the
radiative efficiency, $\dot{M}$ is the black hole accretion rate, and $c$ is
the speed of light. In this calculation, the black holes are assumed to be
non-rotating. If the black holes are spinning, their radiative efficiencies
and luminosities would be higher due to the shrink of the innermost stable
circular orbit, by up to a factor of 4 for maximal rotation.
  
The B-band luminosity of each source is corrected for attenuation by absorption
from the ISM along the line-of-sight. We first calculate the line-of-sight
column-density of the gas from each source to a distant observer. 
For each black hole we generate 1000 radial sight lines originating at the
black hole particle location and uniformly spaced in the $4\pi$ solid angle
${\rm d}cos{\theta}\, {\rm d}\phi$, while for the stars, an accurate estimate
of the luminosity is possible with only one sight line per source owing to the
extended distribution.  Along each ray, the gas column density is
calculated using a radial spacing of $\Delta r=\eta h_{\rm sml}$, where $\eta
\leq 1$ and $h_{\rm sml}$ is the local SPH smoothing length. The distribution
of line-of-sight properties converges for $\gtrsim 100$ rays and at a distance
of $\gtrsim 100\, \kpc$. In the calculation, only the diffuse-phase density is
considered because of its large volume filling factor $\ge 99\%$, allowing
for a determination of the lower limit on the column density along a
particular line of sight.  

Adopting the mean observed intrinsic quasar continuum spectral  energy
distribution \citep{Richards2006} gives a B-band luminosity which is well
approximated by the following equation given by \cite{Marconi2004},
$\log{(\Lbol/\LB)}=0.80-0.067{L}+0.017{L}^{2}-0.0023{L}^{3}$, where
${L} = \log{(\Lbol/\Lsun)} - 12$, and $\lambda_{B}=4400\, \AA$. We then use
the Milky Way gas-to-dust ratio scaled by metallicity, $A_{B}/N_{H} =
(Z/0.02)(A_{B}/N_{H})_{\rm MW}$ to determine the extinction along a given line 
of sight for this band. In the above calculation, we do not include a full
treatment of radiative transfer, and therefore do not model scattering or
re-processing of radiation by dust in the infrared. However, for the B-band
luminosity, results using a 3-D Monte Carlo radiative transfer code are
close to those calculated using the methods we present here
\citep{Li2007A, Chakrabarti2006}.   

Figure~\ref{FigLum} shows both the bolometric and attenuated B-band
luminosities of the quasar, compared with observations of \zquasar. The system
is intrinsically bright with a total luminosity $>10^{11}\, \Lsun$, and the
host appears as an ultraluminous infrared galaxy (ULIRG) with $\Lbol
> 10^{12}\, \Lsun$ for most of the time. At high redshifts, $z>8$, starlight
dominates the total luminosity. However, black holes take over at a later
time. The quasar light-curve increases dramatically, peaking at $z
\approx 6.5$, when it is powered by the most massive black hole accreting at
near the Eddington rate. The estimated $\Lbol$ of \zquasar\ differs from that
of the simulated quasar by less than the uncertainty in the luminosity
estimate. The rest-frame B-band absolute magnitude reaches $\rm M_{\rm B}
\sim -26.5$, almost one magnitude fainter than that of \zquasar\ derived from
1450$\AA$\ data \citep{Fan2003}. However, we should emphasize that in this
paper, our main goal is to investigate the plausibility of forming luminous $z
\sim 6$ quasars through hierarchical mergers, rather than precisely
reproducing the properties of an individual quasar such as \zquasar, so the
disagreement shown in Figure~\ref{FigLum} should not be taken too
literally. Moreover, the {\em exact} luminosity can change by a factor of
several from relatively trivial or random details in the simulations. If the
black hole spin is taken into account, then the simulated luminosities would
increase by a factor of up to 4, which would match the observation of
\zquasar\ better.

Feedback-driven outflows create un-obscured lines-of-sight, allowing the
growing central SMBH to be visible as an optically-bright quasar between
redshifts $z \sim 7.5$ and $z \sim 6.4$. At the peak of the quasar activity,
more than 50\% of the 1000 sight lines have $\LB \ge 10^{12}\, \Lsun$. The
absorbed light is re-emitted at infrared wavelengths by dust. We find that 
the luminosity in the far infrared \citep{Li2007A} is close to $\Lfir \sim
10^{13}\, \Lsun$ estimated for \zquasar\ by \cite{Bertoldi2003A}. Moreover, we
find that up to  80\% of the FIR light comes from the black hole, while stars
contribute only $\sim 20\%$. This may explain why the star formation rate at
$z\approx 6.5$ during the peak quasar phase is an order of magnitude lower
than the estimate from FIR observations \citep{Bertoldi2003A}, which will be
contaminated by the AGN.

Another prominent feature of Figure~\ref{FigLum} is a clear phase transition
from starburst to quasar. It has long been suggested that ULIRGs are powered
by starbursts in galaxy mergers (for reviews, see \citealt{Sanders1996,
Jogee2006}), and that bright quasars are the descendants of ULIRGs
(\citealt{Sanders1988, Norman1988, Scoville2003, Alexander2005}). This
conjecture has been supported by observations of quasar hosts (e.g.,
\citealt{Stockton1978, Heckman1984, Hutchings1992, Bahcall1997,
Hutchings2005}), and theoretical modeling \citep{Hopkins2006A}.  In
\cite{Li2007A}, we calculate the spectral energy distributions (SEDs) of the
quasar system and its galaxy progenitors. We find that the SEDs of the system at
$z>8$ are characterized by those of starburst galaxies \citep{Sanders1996},
while at the peak quasar phase, the SEDs resemble those observed in $z \sim 6$
quasars \citep{Jiang2006}. We also find that the system evolves from cold to
warm ULIRG as it transforms from starburst to quasar phase. Our results
provide further theoretical evidence for the ULIRG--quasar connection in
quasar systems in the early Universe.

The quasar lifetimes depend on the observed luminosity threshold, as proposed
by \cite{Hopkins2005A}. In our simulation, at the peak luminosity of $\LB
\approx 2 \times 10^{12}\, \Lsun$, the quasar lifetime is roughly $\sim 2 \times
10^{6}$ yrs, as shown in Figure~\ref{FigTq}.  Again, If black hole spin is
included in the calculation, the luminosity of the quasar would increase by a
factor of several, and the quasar lifetime would be longer. However, when
increasing the radiative efficiency, the Salpeter time ($e$-folding time for
Eddington-limited black hole growth, \citealt{Salpeter1964}) is 
increased by an identical factor, meaning it would also require a longer time 
to reach the same mass. If high-redshift quasars are rapidly rotating, then,
our calculations demand that either the seeds be {\em much} more massive at
$z \gtrsim 6$, or that they accrete in a super-Eddington manner. In other
words, if the observed Sloan quasars at $z \sim 6$ shine with Eddington
luminosity but are rotating rapidly, then our model suggests that their masses
would be considerably smaller than estimated.

We note that recent {\em Spitzer} observations by \cite{Jiang2006} show that 2
out of 13 quasars at $z\sim 6$ have a remarkably low NIR-to-optical flux ratio
compared to other quasars at different redshifts, and these authors suggest
that the two quasars may have different dust properties. According to our
model, however, these two outliers may be young quasars that have just
experienced their first major starburst but have not yet reached peak quasar
activity, so the light from star formation may be dominant, or comparable to
that from the accreting SMBH still buried in dense gas. This may explain the
low NIR flux, as well as the B-band luminosity and the narrow $L_{\alpha}$
emission line, which are primarily produced by the starburst. We will address
this question further in a future paper with detailed modeling and IR
calculations \citep{Li2007A}.    

\section{Discussion}

\subsection{Comparison with Previous Models and Robustness of Our Results}
Our multi-scale simulations that include large-scale cosmological N-body
calculations and hydrodynamic simulations of galaxy mergers, as well as a
self-regulated model for black hole growth, have successfully produced a
luminous quasar at $z \sim 6.5$ with a black hole mass of $\sim 2
\times10^{9}\, h^{-1}\, \Msun$ and a number of properties similar to those of 
\zquasar, the most distant quasar detected at $z=6.42$ \citep{Fan2003}. Our
approach differs from previous semi-analytic studies by \cite{Haiman2001,
  Haiman2004, Yoo2004, Volonteri2005} and \cite{Volonteri2006} in the
following ways: 

\begin{enumerate}

\item We use a realistic merger tree derived directly from multi-scale,
  high-resolution cosmological simulations. The previous studies used merger
  trees of dark matter halos generated with the extended Press-Schechter
  theory \citep{Press1974, Lacey1993}, which may underestimate the abundance
  of high-mass halos by up to one order of magnitude, as shown in
  \S~\ref{subsec_mf}. Also, the merger trees in those studies started from
  much higher redshifts than what we consider here. In our model,
  the quasar halo is the largest one in a volume of $1\, h^{-3}\,
  \Gpc^{3}$. It has a mass of $\sim 7.7 \times 10^{12}\, \Msun$ at $z \sim
  6.5$ built up through seven major mergers from $z \simeq 14.4$ to $z \simeq
  6.5$. 

\item We follow the evolution of the system and treat the gas dynamics, star
  formation, and black hole growth properly. This approach is
  critical to investigation of the properties of both black holes and
  host galaxies, and their evolution (e.g., \citealt{DiMatteo2005,
  Springel2005B, Robertson2006A, Hopkins2006A}), but it was not included in
  those previous studies on formation of $z \sim 6$ quasars. 

\item We employ a self-regulated model for the growth of supermassive black
  holes, in which the accretion is regulated by the black hole feedback, and
  the rate is under the Eddington limit. In the previous studies, the black
  hole growth was unregulated, but instead a constant- or super-Eddington
  accretion rate was used.  

\item In our simulations, we do not consider black hole ejection caused by
  gravitational recoil owing to insufficient resolution and lack of
  relativistic physics. However, the halo escape velocities in our
  simulations are in the range of $486$ -- $1284\, \rm{km\, s}^{-1}$, much
  larger than the kick velocity $\sim 100$ -- $475\, \rm{km\, s}^{-1}$ (e.g.,
  \citealt{Herrmann2006, Baker2006, Gonzalez2006, Herrmann2007}, see
  \S~\ref{subsec_Vesc} and \S~\ref{subsec_bhgrowth} for more
  details). Therefore, black hole ejection may be negligible in our case.
  Previous studies had much smaller halo progenitors at higher redshifts than
  ours, so the black hole seeds would be more likely subject to ejection from
  their halos. This leads to the conclusion in these studies that constant or
  super-Eddington accretion is needed owing to significant black hole
  ejection. Our results are robust within the best estimates
  currently available for the recoil velocity of the black hole binary. 

\item The black hole seeds in the galaxy progenitors in our simulations are
  massive (e.g., $\sim 10^5\, \Msun$ at $z \sim 14$). The sub-resolution
  recipe in our model does not allow us to resolve the actual formation
  and accretion of such black holes below this mass scale. The formation of
  these seeds is an unsolved problem, but our results do not depend on the
  specific prescription of the formation process. We adopt a picture in which
  the seed holes come from the remnants of the first stars (which have a
  mass $200\, \Msun$ at $z = 30$) and grow under Eddington limit until they
  enter the merger tree we simulated. If the growth is delayed by
  radiation feedback from the PopIII stars (e.g., \citealt{Johnson2006}), then
  super-Eddington accretion, or other proposed scenarios (e.g.,
  \citealt{Bromm2003, Begelman2006}) may be necessary to
  form massive seeds of $\sim 10^{5}\, \Msun$ in the protogalaxies.  
\end{enumerate}

Overall, we conclude that the results from our simulations, which are
more realistic and more detailed than the models previously done, are
robust. Suppermassive black holes of $\sim 10^9\, \Msun$ can form rapidly
through gas-rich hierarchical mergers under Eddington limit, even within a
short period of time. We find that constant or super-Eddington
accretion is not necessary unless the above assumptions in our modeling break,
i.e., there are no massive black hole seeds of $10^5\, \Msun$ available at $z
\sim 14$, or the recoil velocity of the coalescing black hole binary is
extremely high (e.g., $> 1000\, \rm{km\, s}^{-1}$). Under these extreme
circumstances, some ``exotic'' processes such as super-Eddington accretion may
be necessary to grow a $\sim 10^9\, \Msun$ SMBH within a few hundred million
years. However, we should note, as pointed out by \cite{Bogdanovic2007}, that
most gas-rich galaxy mergers have a configuration such that the orbit and
spins of both black holes are aligned with the large-scale gas flow owing to
torques from accreting gas. Such a configuration has a maximum kick velocity
$< 200\, \rm{km\, s}^{-1}$, which is well below the escape velocity of a
$10^{10}\, \Msun$ dwarf galaxy, as well as those of the halos in our modeling.   

\subsection{Merging History of Black Holes}
\label{subsec_bhbinary}

During the galaxy mergers, the black holes follow their host halos to 
the system center and can form binaries (or multiple systems). The coalescence 
of a black hole binary includes three distinct phases: inspiral, merger, and
ringdown (e.g., \citealt{Flanagan1998}). Whether black hole binaries can
coalesce on short timescales is a matter of debate. In a stellar environment, 
it has been argued that a binary hardens very slowly owing to an eventual
depletion of stars that cause the binary to lose angular momentum (e.g.,
\citealt{Begelman1980, Milos2003}). In a gaseous environment, 
however, numerical simulations by \cite{Escala2004} and \cite{Li2007C}
show that the binaries decay rapidly owing to strong dynamical friction with
the gas, and they likely merge within $10^7$ years. Because
our galaxies are very gas rich and have large central concentrations of gas
during the mergers, we assume the black hole particles coalesce once their
separation decreases below our spatial resolution ($30\, h^{-1}\, \rm{pc}$)
and their relative speed falls below the local gas sound speed
\citep{Springel2005B}.    

In the simulations, we do not have sufficient resolution nor the relativistic
physics to consider the ejection of black holes by gravitational recoil during
the merger phase. However, as discussed in \S~\ref{subsec_Vesc} and
\S~\ref{subsec_bhgrowth}, the halo escape velocities in our simulations 
are much larger than the maximum kick velocity for black hole binary estimated
from the latest relativistic calculations. So black hole ejection is likely
unimportant in our modeling. To accurately address gravitational recoil in the
galaxy merger simulations, we need to include general relativity, resolve the
dynamics of black hole binaries with extremely high resolution, and calculate
the halo potential in a cosmological context (in which halo potential
distribution may be different from that of a single object). However, such a
comprehensive treatment is impossible at the moment. We therefore assume the
black holes merge quickly once they reach the stage of gravitational
radiation.    

These coalescing supermassive black holes will be strong sources of
gravitational radiation detectable by the Laser Interferometer Space Antenna
({\em LISA}, \citealt{Folkner1998}), as suggested by many authors (e.g.,
\citealt{Thorne1976, Haehnelt1998B, Flanagan1998, Menou2001, Hughes2002,
  Sesana2005, Koushiappas2006}). By tracing the merging history of 
the SMBHs, {\em LISA} could shed light on the distribution, structures and
evolution of the associated dark matter halos. Because luminous, high-redshift
quasars are likely sites of vigorous hierarchical mergers, they may be the
best targets for {\em LISA} to explore the early Universe. 

\subsection{Feedback from Starburst-driven Winds}

\begin{figure}
\begin{center}
\includegraphics[width=3.5in]{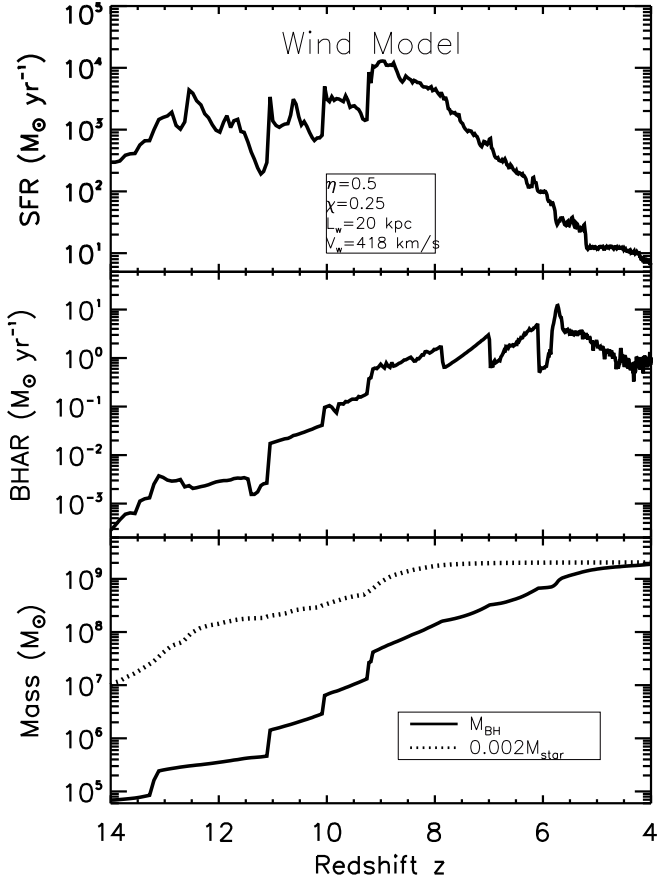}
\vspace{0.5cm}
\caption{Evolution of star formation and black hole growth in merger
  simulations with a starburst-driven wind model. The simulation is
  run with lower resolution ($\rm {N_{tot}} \sim 5\times 10^5$), and the
  specifications of the wind model is: wind efficiency $\eta=0.5$, wind energy
  coefficient $\chi=0.25$, wind free travel length $\rm {L_{w} = 20\, \kpc}$,
  and a wind velocity $\rm {V_{w} = 418\, km\, s^{-1}}$.}  
\label{FigWind}
\end{center}
\end{figure}

Vigorous star formation would induce a galactic wind and mass outflow, a 
phenomenon that has been observed to prevail in both local star-forming
galaxies as indicated by blueshifted optical absorption lines (e.g.,
\citealt{Martin1999, Heckman2000, Martin2005, Rupke2002, Rupke2005}), and
Lyman Break Galaxies at $z \sim 3$ as indicated by blueshifted interstellar
absorption lines and redshifted Ly$\alpha$ emission lines (e.g.,
\citealt{Pettini2002, Shapley2003}), as well as Ly$\alpha$ 
emitters at $z \sim 5.7$ \citep{Ajiki2002}. These galactic winds are generally
thought to play a significant role in galaxy evolution (e.g., see
\citealt{Veilleux2005} and \citealt{Cox2006C} for recent reviews).   

The strong starburst preceding the major quasar phase in our simulations may
drive strong galactic winds and affect the black hole growth. To investigate
the impact of the feedback from a starburst-driven wind on the growth of the
black hole, we have done the same merger simulation with lower resolution
($\rm {N_{tot}} \sim 5\times 10^5$) and with a canonical wind model from 
\cite{Springel2003b}: the wind efficiency $\eta=0.5$, which measures the
coefficient of the star formation that determines the mass outflow; the energy
fraction from supernovae injected into the wind $\chi=0.25$, wind free travel
length $\rm {L_{w} = 20\, \kpc}$, and a wind velocity $\rm {V_{w} = 418\, km\,
  s^{-1}}$. As demonstrated by \cite{Cox2006C}, this wind model is able to
reproduce the starbursts as observed in Lyman Break Galaxies, and therefore is
suitable to our study.  

We find that the impact of the starburst-driven wind on the quasar
evolution is minor, as shown in Figure~\ref{FigWind}. Both the histories of
star formation and black hole growth remain roughly the same as in the
simulation without a starburst wind, only the amplitude is lowered by a factor
of $\sim 1.5$. Similarly, the final masses of the black hole and the
stars are reduced by roughly the same factor, but the quasar host is still on
the $\mbulge$ correlation. The peak quasar phase is delayed to $z \sim 6$.
Overall, the starburst wind affects the gas dynamics locally, but owing to the
deep potential of the system, its impact on the process of quasar
formation is minor. Our results support the finding by \cite{Cox2006C}
that feedback from starburst-driven winds alone is ineffective at regulating
the growth of the central black hole, so feedback from the black hole plays the
dominant role in the formation and evolution of quasars.     

\subsection{Abundance and Fate of Quasar Halos at $z \sim 6$}

\begin{figure}
\begin{center}
\includegraphics[width=3.5in]{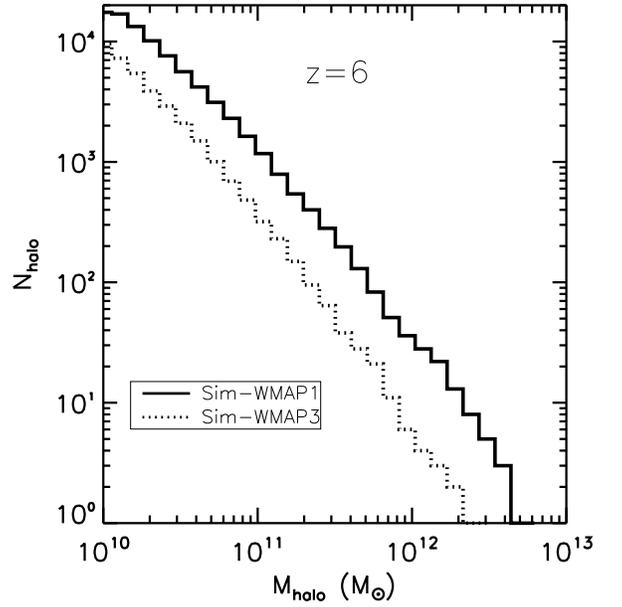}
\caption{Comparison of halo abundances at $z \sim 6$ from the zoom-in
  simulations with parameters from both WMAP1 (solid line) and WMAP3
  (dashed line). The volume of the high-resolution zoom-in region is $\sim
  50^3 \, h^{-3}\, \Mpc^3$.}   
\label{FigNum}
\end{center}
\end{figure}

Because we have so far simulated only one quasar in a volume of 
$1 \, h^{-3}\, \Gpc^{3}$, we are not yet able to
constrain the expected abundance of quasars $z\sim 6$. As
mentioned in \S~2.3, at a given redshift, cosmological simulations with
parameters from WMAP1 produce more massive halos than runs with WMAP3 owing to
a larger value of $\sigma_8$. Figure~\ref{FigNum} shows the number of halos at
$z\sim 6$ from the zoom-in runs with parameters from both WMAP1 and
WMAP3. There are about three dozen halos with mass $M > 10^{12}\, \Msun$ in
the WMAP1 run, while in the WMAP3 run there are only a handful of such
halos. However, since in our picture the quasar activity depends not only on
the halo mass, but also on the merging history, an accurate estimate of the
quasar abundance and luminosity function would require hydrodynamical
simulations of all the quasar candidates in a large box, which are currently
unavailable. Nevertheless, all conditions being equal, the change from the
WMAP3 parameters would produce fewer luminous quasars at $z \sim 6$. This
suggests that in a WMAP3 cosmology, the quasar observed with the largest
redshift, \zquasar, might have formed in a slightly higher overdensity peak
than that we have presented here. In that event, if the WMAP3 determination of
$\sigma_8$ were correct, we would need to identify a rarer density fluctuation
to match \zquasar\ at its observed redshift. However, this does not
change our conclusion that the most distant and luminous quasars can form from
hierarchical galaxy mergers in the $\Lambda$CDM cosmology.  

Imaging surveys of \zquasar\ show that there is no other luminous quasar from the
same epoch in the field (\citealt{Carilli2004, White2005, Willott2005}). In our
simulations, around the peak of quasar activity at $z\approx 6.5$, there are
no other halos of mass $> 10^{12}\, \Msun$ within a few Mpc of this quasar. However,
the numerous major mergers this halo experienced prior to the peak quasar
activity demonstrates that this region was once highly clustered with massive
halos, but they merged to become a bigger one by $z = 6.5$. 

As seen from Figure~\ref{FigTree}, this quasar halo will undergo a handful of
major mergers at a later time from $z\sim$~4--1, and eventually end up as a
cD-like galaxy at the center of a rich cluster. Since we do not follow
hydrodynamically the evolution of the quasar at $z<4$, the physical conditions
of these mergers remain undetermined. It is not clear whether this halo would
experience more episodes of starburst or quasar activity later on during
these mergers. Therefore, the final black hole mass and other properties of
this quasar at the present day are deferred to future simulations that follow its
evolution to $z=0$.

\subsection{Galaxies in the Epoch of Reionization}

The epoch of reionization (EoR) is an important landmark event
in cosmic history that constrains the formation of the first luminous objects
\citep{Loeb2001}. The recent results of WMAP3 indicate that the Universe was 50\%
reionized at $z \approx 9.3$ \citep{Page2006, Spergel2006}, while studies of
Gunn-Peterson absorption \citep{Gunn1965} suggest that reionization began
as early as $z \sim 14$ and ended at $z \sim 6$ \citep{Fan2006A}. At present,
it is believed that the reionization sources are star-forming galaxies since
there are insufficient quasars at $z>6$ as indicated by the steep quasar
luminosity function \citep{Fan2006A}. 

The galaxy progenitors of the quasar in our simulations underwent extreme
and prolonged starbursts before $z \sim 6.5$. Less extreme galaxies in this
epoch may also have vigorous star formation histories. Detecting these
galaxies and determining their contribution to reionization will be crucial
to understanding the EoR (\citealt{Hernquist2003, Barton2004,
Dave2006}). As reviewed by \cite{Hu2006}, recent observations using both
broad-band colors (e.g., \citealt{Dickinson2004, Yan2005, Bunker2004,
Bouwens2004, Giavalisco2004, Egami2005, Eyles2005, Mobasher2005, Yan2006,
Eyles2006}) and narrow-band Ly$\alpha$ emission (e.g., \citealt{Hu2002,
Malhotra2004, Stern2005}) have detected $\sim 500$ galaxies at $z \sim 6$ and
a handful at $z \gtrsim 7$ \citep{Bouwens2005}. The low luminosity density of
galaxies currently detected at $z>7$ seems insufficient to reionize the
Universe. However, ongoing surveys with the HST and Spitzer telescopes, and
future missions such as the {\em Dark Ages z Ly$\alpha$ Explorer} (DAzLE,
\citealt{Horton2004}) and the {\em James Webb Space Telescope} (JWST,
\citealt{Windhorst2006}) will search deeper and further for high-redshift
objects, and may eventually unveil ionizing sources in the EoR.  

\section{Summary}

We have presented a model that accounts for the SMBH growth, quasar activity
and host galaxy properties of the most distant quasar observed at $z = 6.42$,
by following the hierarchical assembly of the quasar halo in the standard
$\Lambda$CDM cosmology.  We employ a set of multi-scale simulations that 
include large-scale cosmological N-body calculations and hydrodynamic
simulations of galaxy mergers, and a recipe for black hole growth
self-regulated by feedback. We first perform a coarse N-body simulation in a
volume of $1\, h^{-3}\, \Gpc^{3}$ to identify the largest halo at
$z=0$, which is assumed to be the descendant
of the earliest luminous quasar. We 
then ``zoom in'' on the halo and resimulate the region with higher
resolution sufficient to extract its merging history starting from very high
redshift. The largest halo at $z \sim 6$ reaches a mass of $\sim 5.4
\times10^{12}\, h^{-1}\, \Msun$ through 7 major mergers between $z \sim$
14.4--6.5. These major mergers are again re-simulated hydrodynamically using 
galaxy models and recipes for star formation, SMBH growth, and feedback.

We find that the quasar host galaxy builds up rapidly through gas-rich
major mergers, with star formation rates up to $10^4\, \Msun\,
\yr^{-1}$, reaching a stellar mass of $\sim 10^{12} \, \Msun$ at $z
\sim 6.5$. The black holes grow through gas accretion under the
Eddington limit in a self-regulated manner owing to their own
feedback. As the galaxies merge, the black holes coalescence to form a
dominant black hole, reaching a peak accretion rate of $\sim 20\, \Msun\,
\yr^{-1}$ and a mass of $\MBH \sim 2 \times 10^9 \, \Msun$ at $z\sim
6.5$. Feedback from black hole accretion clears away the
obscuring gas from the central regions, making the quasar optically
visible from $z \sim$~7.5--6. At the peak of the quasar phase, the star
formation rate, metallicity, black hole mass, as well as quasar
luminosities of the simulated system are consistent with observations
of \zquasar.

Our results demonstrate that rare and luminous quasars at high redshifts can
form in the standard $\Lambda$CDM cosmology through hierarchical, gas-rich
mergers, within the available cosmic time up to the early epoch of $z\approx
6.5$, without requiring exotic processes. Our model should also provide a
viable formation mechanism for other distant, luminous quasars. Moreover, we
predict that quasar hosts at high redshifts follow similar $\mbulge$
correlation observed locally as a result of the coeval evolution of the SMBHs
and host galaxies. Better measurements of black hole masses and host
properties with future observations will therefore be crucial to test our
prediction. Furthermore, we predict that the progenitors of the distant  
quasars undergo strong and prolonged starbursts with rates $\sim 10^3 \, \Msun
\, \yr^{-1}$ at higher redshifts $z>8$, that would contribute to the
reionization of the Universe. Detecting these early galaxies and unveiling the
epoch of reionization will be an important goal of current and future missions
in observational cosmology.    
 
\acknowledgments{We thank Volker Bromm, Romeel Dav{\' e}, Suvendra Dutta,
  Zoltan Haiman, Scott Hughes, Adam Lidz, Chris McKee, Desika Narayanan,
  Hans-Walter Rix, Alice Shapley, Rachel Somerville, Curt Struck, Naoshi
  Sugiyama, Rashid Sunyaev, and Scott Tremaine for stimulating discussions,
  and the referee for helpful comments. YL gratefully acknowledges an
  Institute for Theory and Computation 
  Fellowship, and financial support from the Grants-in-Aid for Young Scientists
  A17684008 by MEXT (Japan) for a visit to Nagoya. BR was supported in part by NASA
  through the Spitzer Space Telescope Fellowship Program through a contract
  issued by the Jet Propulsion Laboratory, California Institute of Technology,
  under a contract with NASA. NY acknowledges support from the Grants-in-Aid
  for Young Scientists A17684008 by MEXT. The computations reported here were
  performed at the Center for Parallel Astrophysical Computing at
  Harvard-Smithsonian Center for Astrophysics, while the initial conditions
  for the cosmological simulations were created on the COSMOLOGY MACHINE
  supercomputer at the Institute for Computational Cosmology in Durham. This
  work was supported in part by NSF grant 03-07690 and NASA ATP grant NAG5-13381.}


\end{document}